\documentclass[longauth]{aa}  
\usepackage{graphicx}
\usepackage{txfonts}
\usepackage{multirow}
\usepackage{hyperref}
\hypersetup{colorlinks=true, citecolor=blue, linkcolor=blue}
\newcommand{\mbh}{{\mbox{$M_\mathrm{BH}$}}}
\newcommand{\hb}{{\mbox{H$\beta$}}}
\newcommand{\brg}{{\mbox{Br$\gamma$}}}
\newcommand{\paa}{{\mbox{Pa$\alpha$}}}
\newcommand{\heiia}{He{\sevenrm\,II}\,$\lambda$4686}
\newcommand{\heii}{He{\sevenrm\,II}}
\newcommand{\hei}{He{\sevenrm\,I}}
\newcommand{\feii}{Fe{\sevenrm\,II}}

\newcommand{\OIIIb}{[O{\sevenrm\,III}]\,$\lambda$5007}
\newcommand{\OIIIc}{[O{\sevenrm\,III}]\,$\lambda\lambda$4959,5007}
\newcommand{\OIII}{[O{\sevenrm\,III}]}
\newcommand{\NI}{[N{\sevenrm\,I}]}
\newcommand{\uas}{\mbox{$\mu$as}}

\newcommand{\kms}{{\mbox{$\mathrm{km\,s^{-1}}$}}}
\newcommand{\uhc}{{\mbox{$\mathrm{km\,s^{-1}~Mpc^{-1}}$}}}
\newcommand{\degree}{{\mbox{$^\circ$}}}
\newcommand{\hi}{H{\sevenrm\,I}}
\font\sevenrm=cmr7 scaled 1000

\begin{document}

\title{A geometric distance to the supermassive black Hole of NGC~3783}

   \authorrunning{GRAVITY Collaboration}
\author{GRAVITY Collaboration\thanks{GRAVITY is developed
    in a collaboration by the Max Planck Institute for
    extraterrestrial Physics, LESIA of Observatoire de Paris/Universit\'e PSL/CNRS/Sorbonne Universit\'e/Universit\'e de Paris and IPAG of Universit\'e Grenoble Alpes /
    CNRS, the Max Planck Institute for Astronomy, the University of
    Cologne, the CENTRA - Centro de Astrofisica e Gravita\c c\~ao, and
    the European Southern Observatory.}:
A.~Amorim\inst{15,17}
\and M.~Baub\"ock\inst{1} 
\and M.~C.~Bentz\inst{23}
\and W.~Brandner\inst{18} 
\and M.~Bolzer\inst{1} 
\and Y.~Cl\'enet\inst{2} 
\and R.~Davies\inst{1}
\and P.~T.~de~Zeeuw\inst{1,13} 
\and J.~Dexter\inst{20,1}
\and A.~Drescher\inst{1,22} 
\and A.~Eckart\inst{3,14} 
\and F.~Eisenhauer\inst{1} 
\and N.~M.~F\"orster~Schreiber\inst{1} 
\and P.~J.~V.~Garcia\inst{11,16,17} 
\and R.~Genzel\inst{1,4} 
\and S.~Gillessen\inst{1} 
\and D.~Gratadour\inst{2,21} 
\and S.~H\"onig\inst{5}
\and D.~Kaltenbrunner\inst{1}
\and M.~Kishimoto\inst{6} 
\and S.~Lacour\inst{2,12} 
\and D.~Lutz\inst{1} 
\and F.~Millour\inst{7}  
\and H.~Netzer\inst{8} 
\and C.~A.~Onken\inst{21}
\and T.~Ott\inst{1} 
\and T.~Paumard\inst{2} 
\and K.~Perraut\inst{9} 
\and G.~Perrin\inst{2} 
\and P.~O.~Petrucci\inst{9} 
\and O.~Pfuhl\inst{12}
\and M.~A.~Prieto\inst{19}  
\and D.~Rouan\inst{2}
\and J.~Shangguan\inst{1}\thanks{Corresponding author: J. Shangguan \newline e-mail: shangguan@mpe.mpg.de}
\and T.~Shimizu\inst{1}
\and J.~Stadler\inst{1}
\and A.~Sternberg\inst{8,10} 
\and O.~Straub\inst{1} 
\and C.~Straubmeier\inst{3} 
\and R.~Street\inst{24}
\and E.~Sturm\inst{1} 
\and L.~J.~Tacconi\inst{1} 
\and K.~R.~W.~Tristram\inst{11}  
\and P.~Vermot\inst{2} 
\and S.~von~Fellenberg\inst{1}
\and F.~Widmann\inst{1} 
\and J.~Woillez\inst{12}}

\institute{
Max Planck Institute for Extraterrestrial Physics (MPE), Giessenbachstr.1, 85748 Garching, Germany
\and LESIA, Observatoire de Paris, Universit\'e PSL, CNRS, Sorbonne Universit\'e, Univ. Paris Diderot, Sorbonne Paris Cit\'e, 5 place Jules Janssen, 92195 Meudon, France
\and I. Institute of Physics, University of Cologne, Z\"ulpicher Stra{\ss}e 77, 50937 Cologne, Germany
\and Departments of Physics and Astronomy, Le Conte Hall, University of California, Berkeley, CA 94720, USA
\and Department of Physics and Astronomy, University of Southampton, Southampton, UK
\and Department of Physics, Kyoto Sangyo University, Kita-ku, Japan
\and Universit\'e C\^ote d'Azur, Observatoire de la C\^ote d'Azur, CNRS, Laboratoire Lagrange, Nice, France
\and School of Physics and Astronomy, Tel Aviv University, Tel Aviv 69978, Israel
\and Univ. Grenoble Alpes, CNRS, IPAG, 38000 Grenoble, France
\and Center for Computational Astrophysics, Flatiron Institute, 162 5th Ave., New York, NY 10010, USA
\and European Southern Observatory, Casilla 19001, Santiago 19, Chile
\and European Southern Observatory, Karl-Schwarzschild-Str. 2, 85748 Garching, Germany
\and Sterrewacht Leiden, Leiden University, Postbus 9513, 2300 RA Leiden, The Netherlands
\and Max Planck Institute for Radio Astronomy, Auf dem H\"ugel 69, 53121 Bonn, Germany
\and Universidade de Lisboa - Faculdade de Ci\^{e}ncias, Campo Grande, 1749-016 Lisboa, Portugal
\and Faculdade de Engenharia, Universidade do Porto, rua Dr. Roberto Frias, 4200-465 Porto, Portugal
\and CENTRA - Centro de Astrof\'isica e Gravita\c{c}\~{a}o, IST, Universidade de Lisboa, 1049-001 Lisboa, Portugal
\and Max Planck Institute for Astronomy, K\"onigstuhl 17, 69117, Heidelberg, Germany
\and Instituto de Astrof\'isica de Canarias (IAC), E-38200 La Laguna, Tenerife, Spain
\and Department of Astrophysical \& Planetary Sciences, JILA, University of Colorado, Duane Physics Bldg., 2000 Colorado Ave, Boulder, CO 80309, USA
\and Research School of Astronomy and Astrophysics, Australian National University, Canberra, ACT 2611, Australia
\and Department of Physics, Technical University Munich, James-Franck-Stra{\ss}e 1, 85748 Garching, Germany
\and Department of Physics and Astronomy, Georgia State University, Atlanta, GA 30303, USA
\and LCOGT, 6740 Cortona Drive, Suite 102, Goleta, CA 93117, USA
}


 
\abstract{
The angular size of the broad line region (BLR) of the nearby active galactic 
nucleus (AGN) NGC~3783 has been spatially resolved by recent observations with 
VLTI/GRAVITY.  A reverberation mapping (RM) campaign has also recently obtained 
high quality light curves and measured the linear size of the BLR in a way that 
is complementary to the GRAVITY measurement.  The size and kinematics of the BLR 
can be better constrained by a joint analysis that combines both GRAVITY and RM 
data. This, in turn, allows us to obtain the mass of the supermassive black 
hole in NGC3783 with an accuracy that is about a factor of two better than that 
inferred from GRAVITY data alone. We derive 
$M_\mathrm{BH}=2.54_{-0.72}^{+0.90}\times 10^7\,M_\odot$.  Finally, and perhaps 
most notably, we are able to measure a geometric distance to NGC~3783 of 
$39.9^{+14.5}_{-11.9}$~Mpc.  We are able to test the robustness of the BLR-based 
geometric distance with measurements based on the Tully-Fisher relation and 
other indirect methods.  We find the geometric distance is 
consistent with other methods within their scatter.  We explore the potential of 
BLR-based geometric distances to directly constrain the Hubble constant, $H_0$, 
and identify differential phase uncertainties as the current dominant limitation 
to the $H_0$ measurement precision for individual sources.
}


   \maketitle

\defcitealias{GCn3783}{GC21}
\defcitealias{GCIRAS}{GC20}

%

\section{Introduction}
\label{sec:intro}

Trigonometry is the basis of distance measurements.  The parallax method uses 
the motion of the Earth around the Sun to measure the angular displacement of a 
nearby star \citep{Bessel1838MNRAS}.  From the Hipparcos satellite of the 
European Space Agency (ESA; \citealt{ESA1997}) to the recent Gaia mission 
\citep{Gaia2016AA}, the parallaxes of more than 1 billion stars in the Milky Way 
have now been measured \citep{Gaia2018AA}. 
The geometric method can also be applied to any object, as long as both its 
physical size ($R$) and angular size ($\Theta$) are measurable, as $D=R/\Theta$.  For a distant extragalactic target, the measured geometric 
distance is the angular diameter distance ($D_A$), including cosmological 
expansion.  In contrast to ``standard candles,'' such as pulsating stars 
\citep[e.g.,][]{Bhardwaj2020JApA} and Type Ia supernovae (SNe~Ia; e.g., 
\citealt{Phillips1993ApJ,Riess1996ApJ}), which measure the luminosity distance 
($D_L$), the geometric method does not rely on the calibration based on the 
so-called distance ladder \citep[e.g.,][]{Riess2009ApJ,Riess2021ApJ}.  

However, objects to which the geometric method can be applied are usually rare 
as it is difficult to measure both $R$ and $\Theta$ for the same target.  The 
distance to the supermassive black hole (BH) at the center of the Milky Way is 
measured to a $<0.5\%$ uncertainty level, based on the 27-year astrometric and 
spectroscopic monitoring of the 16-year orbital motion of the star S2 
\citep{GCdist}.  For detached eclipsing binary stars, the linear size of each 
component can be measured from photometric 
and spectroscopic monitoring, while the angular size of each star can be derived 
from its empirical relation with the color of the star \citep{Lacy1977ApJ}.  
This method has been used to measure the distance to the Large Magellanic Cloud with 
high accuracy \citep{Pietrzynski2019Natur}.  By monitoring the Keplerian motion 
of water maser-emitting gas with very long baseline interferometry (VLBI) 
observations, one can measure the geometric distance to nearby Type-2 Seyfert 
galaxies that host an observable megamaser disk 
\citep{Herrnstein1999Natur,Braatz2010ApJ}.  Baryonic acoustic 
oscillations can also provide $D_A$ using the clustering of galaxies at a  
certain redshift range \citep[e.g.,][]{Eisenstein2005ApJ,Anderson2014MNRAS}.
The broad line region (BLR) of active 
galactic nuclei (AGNs) has also been proposed as a probe of geometric distance 
\citep{Elvis2002ApJ}.  The linear size of the BLR can be measured by the 
reverberation mapping (RM) technique \citep{Peterson2014SSRv}, while the angular 
size of the BLR can be measured by near-infrared (NIR) interferometry 
\citep{Petrov2001CRPhy,Woillez2004SPIE}.  Unlike detached 
eclipsing binaries and megamaser systems, which are difficult to discover, 
AGNs are luminous sources that are commonly found locally as well as out to high 
redshifts, even at $z>6$.  A similar approach has been applied to NGC~4151 by 
resolving the hot dust continuum emission instead of the BLR 
\citep{Honig2014Natur}.

Recently, GRAVITY, the second generation Very Large Telescope Interferometer 
(VLTI) instrument, spatially resolved the BLR of 3C~273 
for the first time using the spectroastrometry (SA) technique 
\citep{Bailey1998SPIE}.  GRAVITY combines all four of the 8 m Unit Telescope 
(UT) beams to yield six simultaneous baselines \citep{GC2017FL}.  
\cite{GC2018Natur} reported the mean radius of the BLR as $46 \pm 10\, \uas$ 
and a BH mass of $(2.6 \pm 1.1)\, \times 10^8\,M_\odot$, which is fully consistent 
with that measured by \cite{Zhang2019ApJ} using 10~year RM data.  
\cite{Wang2020NatAs} conducted the first joint analysis (hereafter, SARM, as the 
combination of SA and RM) and derived an angular diameter distance for 3C~273 of 
$D_A = 551.5^{+97.3}_{-78.7}$~Mpc, corresponding to a Hubble constant 
$H_0=71.5^{+11.9}_{-10.6}\,\uhc$.

The Hubble constant has been measured to an accuracy of a few percent in the 
low-$z$ Universe with various methods.  Using the distance ladder, the SH0ES 
(supernovae H0 for the equation of state) project recently derived 
$H_0=73.5 \pm 1.4\,\uhc$ \citep{Reid2019ApJ}.  This value is consistent with 
other measurements independent of the distance ladder, such as megamaser 
observations \citep[e.g.,][]{Pesce2020ApJ} and observations of 
gravitationally lensed quasars \citep[e.g.,][]{Wong2020MNRAS} based on the 
so-called time-delay distance.  In comparison, $H_0$ inferred from conditions in 
the early Universe tends to be substantially smaller than the low-$z$ values 
(4--6 $\sigma$ significance level, \citealt{Riess2020NatRP}; however, see 
\citealt{Freedman2019ApJ}).  For instance, \cite{Planck2020AAb} 
derive $67.4 \pm 0.5\,\uhc$ based on cosmic microwave background 
observations. 
The high luminosity and remarkably uniform spectral properties of AGNs 
have motivated many attempts to use them as standard candles 
\citep{Baldwin1977ApJ,Collier1999MNRAS,Elvis2002ApJ,Watson2011ApJ,Wang2013PhRvL,Honig2014ApJ,LaFranca2014ApJ,Risaliti2015ApJ}.
In particular, the good correlation between the BLR radius and continuum 
luminosity \citep[e.g.,][]{Kaspi2000ApJ,Bentz2013ApJ} makes it promising to probe 
the $D_L$ based on the RM measurements 
\citep[e.g.,][]{Watson2011ApJ,Wang2013PhRvL}.  However, recent studies reveal 
significant deviations from the R--L relation, primarily driven by the increase 
in the Eddington ratio (\citealt{Du2016ApJa,Du2018ApJb,Du2019ApJ,MartinezAldama2019ApJ,DallaBonta2020ApJ}; 
however, see \citealt{Grier2017ApJa,FonsecaAlvarez2020ApJ}).
It is therefore of great importance to explore the power of the SARM method, 
namely to study the BLR structure in detail and to measure the BLR-based 
geometric distance with the goal to independently test the $H_0$ tension in the 
future.

With $z\approx 0.009730$ \citep{Theureau1998AAS} and a much shorter time 
lag than that of 3C~273, NGC~3783 provides the best opportunity to compare the 
geometric distance derived with SARM to other independent measurements.  A new 
RM campaign of NGC~3783 was reported by \cite{Bentz2021ApJ} with a time lag of 
about 10~days, consistent with previous measurements \citep{Onken2002ApJ}.  
\citet[][hereafter, \citetalias{GCn3783}]{GCn3783} reported a BLR mean angular 
radius of about 70~\uas, and the RM-measured time lag can be reproduced with the 
measured continuum light curve and the best-fit BLR model inferred only 
from GRAVITY data at an assumed $D_A=38.5$~Mpc.  In this paper, we construct a 
Bayesian model to fit the GRAVITY and RM data simultaneously 
(Section~\ref{sec:join}).  The inferred BLR model is entirely consistent with 
the results of \citetalias{GCn3783} 
(Section~\ref{sec:infe}).  We find that the inferred $D_A$ of NGC~3783 is fully 
consistent with distances measured using the Tully-Fisher relation and other 
indirect methods (Section~\ref{sec:dist}).  The application of the SARM method to 
$H_0$ is discussed in Section~\ref{sec:h0}, together with improvements that may 
come with the ongoing upgrade of the GRAVITY instrument. 
This work adopts the following fiducial parameters for a $\Lambda$CDM cosmology: 
$\Omega_m = 0.3$, $\Omega_\Lambda = 0.7$, and $H_{0}=70$ km s$^{-1}$ Mpc$^{-1}$, 
unless otherwise specified.

\section{GRAVITY interferometer and reverberation mapping observations}
\label{sec:data}

GRAVITY interferometric data were collected over 3~years from 2018 to 2020 
through a series of Open Time programs and an ESO Large 
Programme\footnote{Observations were made using the ESO Telescopes at the 
La Silla Paranal Observatory, program IDs 0100.B-0582, 0101.B-0255, 0102.B-0667, 
2102.B-5053, and 1103.B-0626.} with the aim to measure the size of the BLR and 
the mass of the central BH.  Details of the data reduction and analysis can be 
found in \citetalias{GCn3783}.  We only briefly summarize the main points here.  
Phase referenced with its hot dust continuum emission, NGC~3783 was observed 
with MEDIUM spectral resolution ($R\approx 500$) in the science channel and 
combined polarization.  After the pipeline reduction, we exclude data with 
poorer performance in terms of phase reference, keeping those with fringe 
tracking ratio $> 80\%$.  We find that the \brg\ profiles in the GRAVITY spectra 
are consistent considering the calibration uncertainty, which indicates that the 
BLR did not change significantly between observations.  We therefore average the 
differential phase of all of the data into three epochs according to their $uv$ 
coordinates.  The differential phase signal of the BLR is dominated by the 
so-called continuum phase, indicating that the center of the BLR is offset from 
the photocenter of the hot dust continuum emission.  The reconstructed image of 
the NGC~3783 continuum emission displays a secondary component, which is the 
main cause of the continuum phase signal, although the asymmetry of the primary 
continuum emission can also contribute to the continuum phase (\citealt{GCIRAS}, 
hereafter, \citetalias{GCIRAS}).  Following \citetalias{GCn3783}, for our 
analysis we adopt the differential phase data after subtracting the continuum 
phase that is calculated from the reconstructed continuum image.  There could be 
still some residual continuum phase due to the uncertainty of the spatial origin 
of the image reconstruction, so we still allow the BLR to be offset from the 
continuum reference center in the modeling (Section~\ref{sec:blr}).  The 
amplitude of the differential phase due to BLR rotation is about 0.2\degree\ 
(Figure 8(b) of \citetalias{GCn3783}). 

Following \citetalias{GCn3783}, for the model inference we adopt the \brg\ 
profile from our $R\approx 4000$ adaptive optics observations obtained on 20 
April 2019 using SINFONI \citep{Eisenhauer2003SPIE,Bonnet2004SPIE}.  The nuclear 
spectrum is extracted using a circular aperture centered on the peak of the 
continuum and a diameter of about 0.1 arcsec, which is larger than the field of 
view of GRAVITY $\sim 60$~milliarcsec.  This results in higher narrow \brg\ 
emission in the SINFONI spectrum; meanwhile, the broad \brg\ profile from the 
SINFONI spectrum is consistent with those from GRAVITY observations.  The high 
spectral resolution of the former guarantees a robust decomposition of the 
narrow-line component, while the latter suffer from low spectral resolution and 
systematic uncertainties due to the calibration.  Therefore, we prefer to use 
the SINFONI data in the BLR modeling.

The RM data were collected in the first half of 2020.  The details of the 
observations and analyses are reported in \cite{Bentz2021ApJ}.  Briefly, the 
photometric and spectroscopic monitoring was conducted with the Las Cumbres 
Observatory global telescope (LCOGT) network.  A total of 209 $V$-band images 
and 50 spectra were obtained.  The \OIIIc\ doublet region was used to calibrate 
the relative flux of the spectra, resulting in a $\sim 2\%$ accuracy for
relative spectrophotometry of \hb\ throughout the monitoring.  The emission 
line light curves were derived from the calibrated spectra by integrating the 
emission lines with the local continuum fitted and subtracted.  The continuum 
light curve at rest-frame 5100~\AA\ is also measured from the spectra and it is 
used to cross-calibrate the $V$-band light curve.  Both of them are merged into 
the final continuum light curve.  The RMS spectrum shows that \hb, \heiia, 
H$\gamma$, and H$\delta$ are variable.  A time lag of about 10~days is derived 
from the well calibrated light curves of the continuum and H$\beta$ line.  

Previous studies have found that the higher order Balmer lines tend to show shorter 
time lags \citep[e.g.,][]{Kaspi2000ApJ,Bentz2010ApJ}.  High-ionization lines such as 
\heiia\ also show smaller lags than that of low-ionization lines such as Balmer lines  
\citep[e.g.,][]{Clavel1991ApJ,Peterson1999ApJ,Bentz2010ApJ,Grier2013ApJ,Fausnaugh2017ApJ,Williams2020ApJ}.
Photoionization models 
\citep{Netzer1975MNRAS,Rees1989ApJ,Baldwin1995ApJ,Korista2004ApJ} provide 
physical explanations for the radial stratification of the BLR.  The higher order 
Balmer lines, with lower optical depth, are expected to be more efficiently 
emitted by gas with higher density and closer to the central BH, resulting in a 
shorter time lag and higher ``responsivity'' \citep[see][]{Korista2004ApJ}.  
However, the optical depth of \hb\ is $\gtrsim 10^3$ based on typical BLR models 
\citep[e.g.,][]{Netzer2020MNRAS}.  It is very challenging for photoionization 
models such as CLOUDY \citep{Ferland2017RMxAA} to theoretically calculate the 
Hydrogen line emission.

Our joint analysis assumes that RM and GRAVITY probe the same regions of the BLR, which 
is encouraged by the consistent size measured from the \hb\ time lag and GRAVITY 
measurements of the \brg\ line.  \cite{Wang2020NatAs} conducted a joint analysis 
using the \hb\ light curve and GRAVITY measurements of the \paa\ line.  The choice is 
mainly limited by the available measurements, although \cite{Zhang2019ApJ} show 
that the time lags of \hb\ and H$\gamma$ for 3C~273 are consistent.  
It is still unclear however how well the BLR sizes of Hydrogen lines in the optical and 
NIR are consistent with each other.
Unfortunately, the H$\gamma$ and H$\delta$ light curves of NGC~3783 are not 
robustly calibrated, because they are far from the reference \OIII\ lines.  
Their time lags are 2--4 times smaller than that of \hb, while their FWHMs are 
also smaller than that of \hb.  This strongly suggests that the lags of 
H$\gamma$ and H$\delta$ are not physically robust.  \cite{Bentz2021ApJ} also 
report a $\sim 2$-$\sigma$ detection of the lag of \heii\ about 2~days; however, 
the discrepancy of BLR sizes between \hb\ and \heii\ lines are not unexpected. 
We therefore only adopt the \hb\ light curve together with the continuum light 
curve in our joint analysis.  One of our main goals in this work is to test 
whether \hb\ and \brg\ BLR radii are consistent by comparing our geometric 
distance with other independent distance measurements.

\section{SARM joint analysis}\label{sec:join}

\subsection{BLR model and spectroastrometry}\label{sec:blr}

Our BLR model has been introduced in detail by \cite{GC2018Natur} and 
\citetalias{GCIRAS}.  Here, we only provide a brief description of the model 
that is necessary for this work.  The model was first developed by 
\cite{Pancoast2014MNRASa} with the original purpose to model velocity 
resolved RM data.  The BLR is assumed to consist of a large number of 
non-interacting clouds, whose motion is governed only by the gravity of a central 
BH with mass $M_\mathrm{BH}$.  A shifted gamma distribution, 
$r = R_\mathrm{S} + F\, R_\mathrm{BLR} + g\, (1-F)\, \beta^2\, R_\mathrm{BLR}$,
is used to describe the radial distribution of the clouds, where $R_\mathrm{S}$ 
is the Schwarzschild radius, $g=p(x|1/\beta^2, 1)$ is drawn randomly from a 
Gamma distribution, $p(x|a,b) = x^{a-1}e^{-x/b}/(\Gamma(a)\,b^a)$, and 
$\Gamma(a)$ is the gamma function.  
The shape parameter, $\beta$, controls the radial profile to be Gaussian 
($0<\beta<1$), exponential ($\beta=1$), or heavy-tailed ($1<\beta<2$).  The 
weighted mean cloud radius, $R_\mathrm{BLR}$, and fractional inner radius, 
$F=R_\mathrm{in}/R_\mathrm{BLR}$, are fitted.  Clouds are then randomly 
distributed in a disk with the angular thickness $\theta_0$, which ranges from 
$0^\circ$ (thin disk) to $90^\circ$ (sphere).  $\gamma$ controls the 
concentration of the cloud distribution toward the edge of the disk (Equation~(4) of 
\citetalias{GCIRAS}).  The weight of the emission from the clouds is controlled by $\kappa$.  With 
respect to the observer, the near side clouds have higher weight if $\kappa>0$, 
while the far side clouds have higher weight otherwise (Equation~(5) of 
\citetalias{GCIRAS}).  The fractional difference in the number of clouds above and 
below the mid-plane is controlled by $\xi$.  Setting $\xi=0$ means there are equal 
numbers of clouds each side of the mid-plane, while there are no clouds below the mid-plane if 
$\xi=1$.  

We define the velocity of the clouds according to their position and the BH mass.  We draw 
cloud velocities from the parameter space distribution of radial and tangential 
velocities centered around either the circular orbits for bound clouds or 
around the escape velocity for inflowing or outflowing clouds (Equation~(6) 
of \citetalias{GCIRAS}).  The fraction of clouds in bound elliptical orbits 
is controlled by $f_\mathrm{ellip}$.  Clouds in radial orbits (inflowing or 
outflowing) are allowed to be mostly bound, mainly controlled by $\theta_e$.  
A random distribution is drawn with Gaussian dispersion along and 
perpendicular to the ellipse connecting circular orbit velocity and radial 
escape velocity.  Whether the radial motion of the cloud ensemble is inflowing or outflowing 
is controlled by a single parameter $f_\mathrm{flow}$, $<0.5$ for inflowing 
and $>0.5$ for outflowing.  The model further considers a line-of-sight velocity 
dispersion to model the macroturbulence.  As in \citetalias{GCn3783}, we find 
the dispersion parameters of the Gaussian distributions in the phase space are 
not crucial to fit the data.  We fix them to zero, so that the BLR model is the 
same as that adopted in \citetalias{GCn3783}. 

The BLR model is rotated with inclination angle $i$, ranging from $0^\circ$ 
(face-on) to $90^\circ$ (edge-on), and position angle PA.  Line-of-sight 
velocities of the clouds account for the full relativistic Doppler effect and 
gravitational redshift.  The flux of each spectral channel is calculated by 
summing the weights of clouds in each velocity bin.  The model line profile is 
scaled according to the maximum, $f_\mathrm{peak}$.  The photocenter of each 
channel is the weighted average position of all clouds in each bin.  The 
differential phase at wavelength $\lambda$ is calculated as 
\begin{equation}\label{eq:dphi}
\Delta \phi_\lambda = -2\pi \frac{f_\lambda}{1+f_\lambda} \vec{u} \cdot 
(\vec{x}_\mathrm{BLR,\lambda} - \vec{x}_c),
\end{equation}
where $f_\lambda$ is the line flux at wavelength $\lambda$ to a continuum level 
of unity, $\vec{u}$ is the $uv$ coordinate of the baseline, and $\vec{x}_c$ is 
the offset of BLR center from the reference center of the continuum emission, 
which can introduce the continuum phase \citepalias{GCIRAS}.

\subsection{Light curve modeling}\label{sec:lc}

In order to generate the model \hb\ light curve, we need to calculate the 
continuum flux reverberated by clouds with different time lag at a given observed 
time.  We need, therefore, to interpolate and extrapolate the continuum light curve 
taking into account the measurement uncertainty.
The variability of an AGN can be described by a damped random walk model 
\citep{Kelly2009ApJ} for which the covariance function between any two times $t_1$ 
and $t_2$ is
\begin{equation}\label{eq:drw}
S(t_1, t_2) = \sigma_\mathrm{d}^2 \exp \left(-\frac{|t_1 - t_2|}{\tau_\mathrm{d}} \right), 
\end{equation}
where $\sigma_d$ is the long-term standard deviation and $\tau_d$ is the typical 
correlated timescale of the continuum light curve.  We use a Gaussian process to 
model the continuum light curve (e.g., 
\citealt{Pancoast2011ApJ,Pancoast2014MNRASa,Li2018ApJ}).  In the fitting, we 
adopt $\tau_d$ and 
$\hat{\sigma}_\mathrm{d}=\sigma_\mathrm{d}/\sqrt{\tau_\mathrm{d}}$ in order to 
relax the correlation between the two parameters.

We can calculate the model \hb\ light curve by convolving the model continuum 
light curve $l_c(t)$ (Appendix~\ref{apd:lc}) with a so-called transfer function 
$\Psi(\tau)$,
\begin{equation}\label{eq:lhb}
\Tilde{l}_\mathrm{H\beta}(t) = A \int \Psi(\tau) \, l_c(t - \tau) \, d\tau,
\end{equation}
where $A$ is a scaling factor.  The transfer function is the normalized 
distribution of the time lag taking into account the weight of clouds of the 
BLR.

\subsection{Bayesian inference}\label{sec:bi}

Following \citetalias{GCIRAS}, we fit the observed \brg\ profile and differential 
phase with the likelihood function of GRAVITY spectroastrometry, 
\begin{align}\label{eq:lsa}
\mathcal{L}_\mathrm{SA} &= \prod_{i=1} \frac{1}{\sqrt{2\pi}\sigma_{f,i}} \exp 
\left( -\frac{(f_i - \Tilde{f}_i)^2}{2\sigma_{f,i}^2} \right)  \nonumber\\
&\times \prod_{i=1} \frac{1}{\sqrt{2\pi}\sigma_{\phi,i}} \exp 
\left( -\frac{(\phi_i - \Tilde{\phi}_i)^2}{2\sigma_{\phi,i}^2} \right),
\end{align}
where $f_i$ and $\Tilde{f}_i$ are the observed and model fluxes of the \brg\ 
profile; $\phi_i$ and $\Tilde{\phi}_i$ are the observed and model differential 
phases; $\sigma_{f,i}$ and $\sigma_{\phi,i}$ are the measurement uncertainties 
of $f_i$ and $\phi_i$ respectively; and $i$ denotes the $i$th channel.  The 
measured \hb\ light curve is fitted with the likelihood function of RM,
\begin{equation}\label{eq:lrm}
\mathcal{L}_\mathrm{RM} = \prod_{i=1} \frac{1}{\sqrt{2\pi} \sigma_{\mathrm{H\beta},i}} 
\exp \left(-\frac{(l_{\mathrm{H\beta}, i} - \Tilde{l}_{\mathrm{H\beta}, i})^2}{2 \sigma_{\mathrm{H\beta}, i}^2}\right),
\end{equation}
where $l_{\mathrm{H\beta},i}$ and $\sigma_{\mathrm{H\beta},i}$ are the $i$th 
measurements of \hb\ flux and uncertainty.  Therefore, the joint likelihood 
function is the multiplication of Equation~(\ref{eq:lsa}) and 
Equation~(\ref{eq:lrm}), 
$\mathcal{L}_\mathrm{SARM} = \mathcal{L}_\mathrm{SA} \times \mathcal{L}_\mathrm{RM}$.

The physical parameters of the BLR model are the primary parameters to be 
inferred from the joint analysis.  Parameters of the continuum light curve model 
($\vec{x_s}$, $x_q$, $\tau_\mathrm{d}$, and $\hat{\sigma}_\mathrm{d}$) are also 
involved in the fitting.  $\vec{x_s}$ and $x_q$ are the deviation of the 
continuum light curve fluxes and their long-term average from the 
\textit{maximum a posteriori} conditioned by the observed light curve using the 
Gaussian process regression (Appendix~\ref{apd:lc}).  Moreover, $\vec{x_s}$ 
consists of 200 parameters because it is important to densely sample the model 
continuum light curve.  It is however worth emphasizing that $\vec{x_s}$ is well 
constrained by the prior information of the observed continuum data and it only 
enters the likelihood function via Equation~(\ref{eq:lhb}).  We tested the 
fitting using 100--300 points for $\vec{x_s}$.  While there is no significant 
difference in the fitting results, we find that the reconstructed continuum 
light curve with 100 points does not capture some features of the 
observed data in some densely sampled region.  The goodness of the fitting does 
not improve when we adopt 300 points.  Therefore, for the sake of computation 
power, we adopted 200 points in our analysis.  With 221 free parameters in 
total, it is very challenging to sample the parameter space.  We 
utilized the diffusive nested sampling code \texttt{CDNest} \citep{Li2018zenodo} 
to do so.  Diffusive nested sampling \citep{Brewer2011} has 
been shown to be effective for fitting RM data with a BLR model (e.g., 
\citealt{Pancoast2014MNRASa,Pancoast2014MNRASb,Li2018ApJ}), as well as for the joint 
analysis of 3C~273 \citep{Wang2020NatAs}.

We find that $\hat{\sigma}_\mathrm{d}$ and $\tau_\mathrm{d}$ are easily biased 
in the joint fitting if no informative prior is used.  The covariance model 
(Equation~(\ref{eq:drw})) should be able to describe the continuum light curve 
well independently.  Therefore, we opted to optimize the likelihood function of 
the continuum light curve data given the covariance model to constrain 
$\hat{\sigma}_\mathrm{d}$ and $\tau_\mathrm{d}$ in advance of the joint fitting, 
\begin{equation}
\mathcal{L}_c = \frac{1}{\sqrt{(2\pi)^m |\vec{C_{11}}|}} \times \exp 
\left( -\frac{(\vec{l_c} - \vec{E}q)^T \vec{C_{11}}^{-1} (\vec{l_c} - \vec{E}q)}{2} \right),
\end{equation}
where $\vec{C_{11}}=S(\vec{t_1}, \vec{t_1}) + \vec{\sigma_c}^2 \vec{I}$ is the 
model covariance matrix of the measured continuum light curve ($\vec{l_c}$) and 
uncertainties ($\vec{\sigma_c}$); and the scalar $q$ is the long-term average of 
the light curve (see Appendix~\ref{apd:lc} for more details).  The flat priors 
with $(-2, 2)$ and $(-1, 3)$ for $\log\,(\tau_\mathrm{d}/\mathrm{day})$ and 
$\log\,(\hat{\sigma}_\mathrm{d}/\sqrt{\mathrm{day}})$ are wide enough for this 
purpose.  We are able to obtain good constraints of  
$\log\,(\tau_\mathrm{d}/\mathrm{day})=1.75^{+0.87}_{-0.37}$ and 
$\log\,(\hat{\sigma}_\mathrm{d}/\sqrt{\mathrm{day}})=-0.75^{+0.03}_{-0.03}$.  
Therefore, we adopted the priors 
$\mathrm{Norm(1.75, 0.5^2)}$ and $\mathrm{Norm(-0.75, 0.03^2)}$ for 
$\tau_\mathrm{d}$ and $\hat{\sigma}_\mathrm{d}$, respectively in logarithmic 
scale in the joint fitting.  Unlike 3C~273 \citep{Wang2020NatAs,Li2020ApJ}, 
we do not find a long-term trend in the light curve of NGC~3783.  Therefore, 
a simple constant $q$ is enough to describe the long-term average of the light 
curve.  We emphasize that the parameters that we are mainly interested in are 
the parameters of the BLR model as well as $D_A$, while the parameters of 
the light curves are considered as nuisance parameters.

\section{BLR model inference}\label{sec:infe}

\begin{table}
\centering
\renewcommand{\arraystretch}{1.5}
\begin{tabular}{l | r }
\hline\hline
                          Parameters    &  Model Inference              \\ \hline
$\Theta_\mathrm{BLR}\,(\mu\mathrm{as})$ & $71_{-14}^{+22}$              \\ 
$R_\mathrm{BLR}\,(\mathrm{ld})$         & $16.2_{-1.8}^{+2.8}$          \\ 
$D_A$ (Mpc)                             & $39.9_{-11.9}^{+14.5}$        \\
$\log\,(\mbh/M_\odot)$                  & $7.40_{-0.15}^{+0.13}$        \\
\hline\hline
\end{tabular}
\caption{Inferred median of posterior sample and central 68\% credible 
interval for the key parameters of the analysis.  $\Theta_\mathrm{BLR}$ is the 
angular radius of the BLR, which is mainly constrained by the differential phase 
of GRAVITY data.  $R_\mathrm{BLR}$ is the physical radius of the BLR, which is 
primarily constrained by RM data.  $D_A$ is the angular diameter distance.  \mbh\ 
is the BH mass.
}
\label{tab:pk}
\end{table}

The inferred model parameters from the joint analysis are entirely consistent 
with what we have obtained from fitting GRAVITY data only \citepalias{GCn3783}.  
Table~\ref{tab:pk} presents the key parameters of this work, while the full 
model parameters are reported in Table~\ref{tab:par}.  All of the parameters are 
consistent within their 2-$\sigma$ uncertainty levels.  One difference in our 
approach here is that, instead of calculating the \textit{maximum a posteriori} 
as \citetalias{GCn3783}, we simply report the median of the marginalized 
posterior distribution, because the dimension of the parameter space ($>200$) is 
prohibitively high to robustly calculate the \textit{maximum a posteriori}.  
Nevertheless, we confirm that the median and \textit{maximum a posteriori} of 
the posterior sampling using only GRAVITY data always show difference within 
1-$\sigma$.  The fitting results are discussed in detail in 
Appendix~\ref{apd:fit}.  We would like to highlight that the uncertainties of 
several key parameters, in particular \mbh, are reduced by a factor of $\sim 2$ 
after adding the RM light curves in the fitting, which indicates that RM data 
help to constrain the model in a consistent manner with GRAVITY data (see also
\citealt{Wang2020NatAs}).  Our inferred BH mass, 
$2.54_{-0.72}^{+0.90}\times 10^7\,M_\odot$, is fully consistent with the value 
based on RM-only data by \cite{Bentz2021ApJ}.  Our uncertainty is about a factor 
of two larger than their statistical uncertainty; but one should bear in 
mind that this does not include the $\sim 0.3$~dex uncertainty due to the virial 
factor, which is the primary uncertainty of integrated RM 
\citep[e.g.,][]{Ho2014ApJ,Batiste2017ApJ}.  Our best-fit model favors moderate 
inflow.  This is expected because both the GRAVITY data and RM data indicate a 
preference for gas inflow individually.  \cite{Bentz2021ApJ} derived the time 
lags of \hb\ in 5 velocity bins.  The time lag profile across the line is 
slightly asymmetric with the longest wavelength showing the shortest time lag, 
indicating that the gas motion of the BLR is a combination of rotation and 
inflow (however, see \citealt{Mangham2019MNRAS}).

\begin{figure}
\centering
\includegraphics[width=0.45\textwidth]{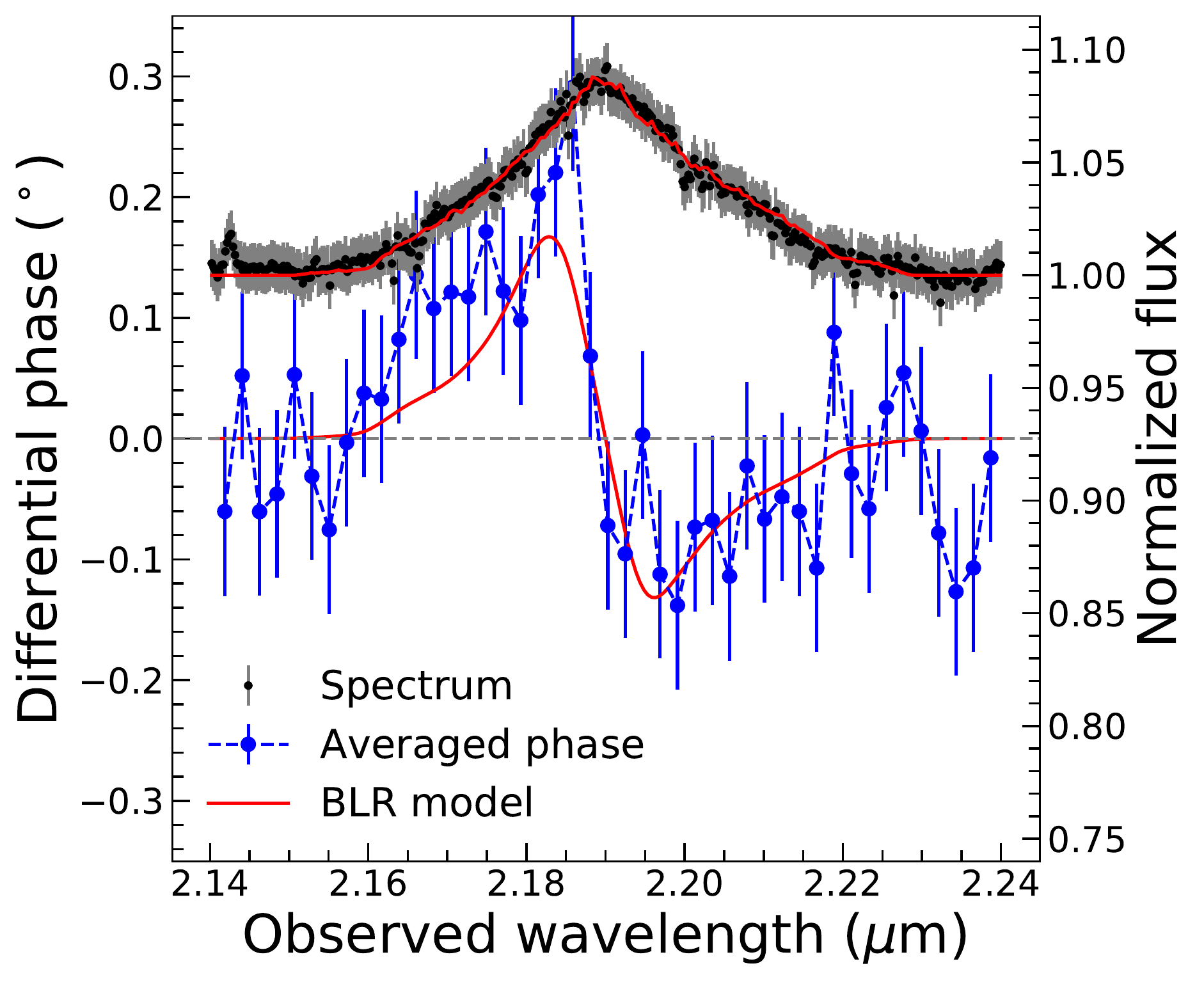}
\caption{Best-fit line profile and differential phase signal of the BLR are 
displayed with red curves.  The \brg\ profile from SINFONI spectrum is in black. 
The averaged differential phase data of UT4--UT2, UT4--UT1, and UT3-UT1 (see 
Figure~\ref{fig:phson}) with the continuum phase subtracted are displayed in 
blue.}
\label{fig:phs}
\end{figure}

\begin{figure}
\centering
\includegraphics[width=0.45\textwidth]{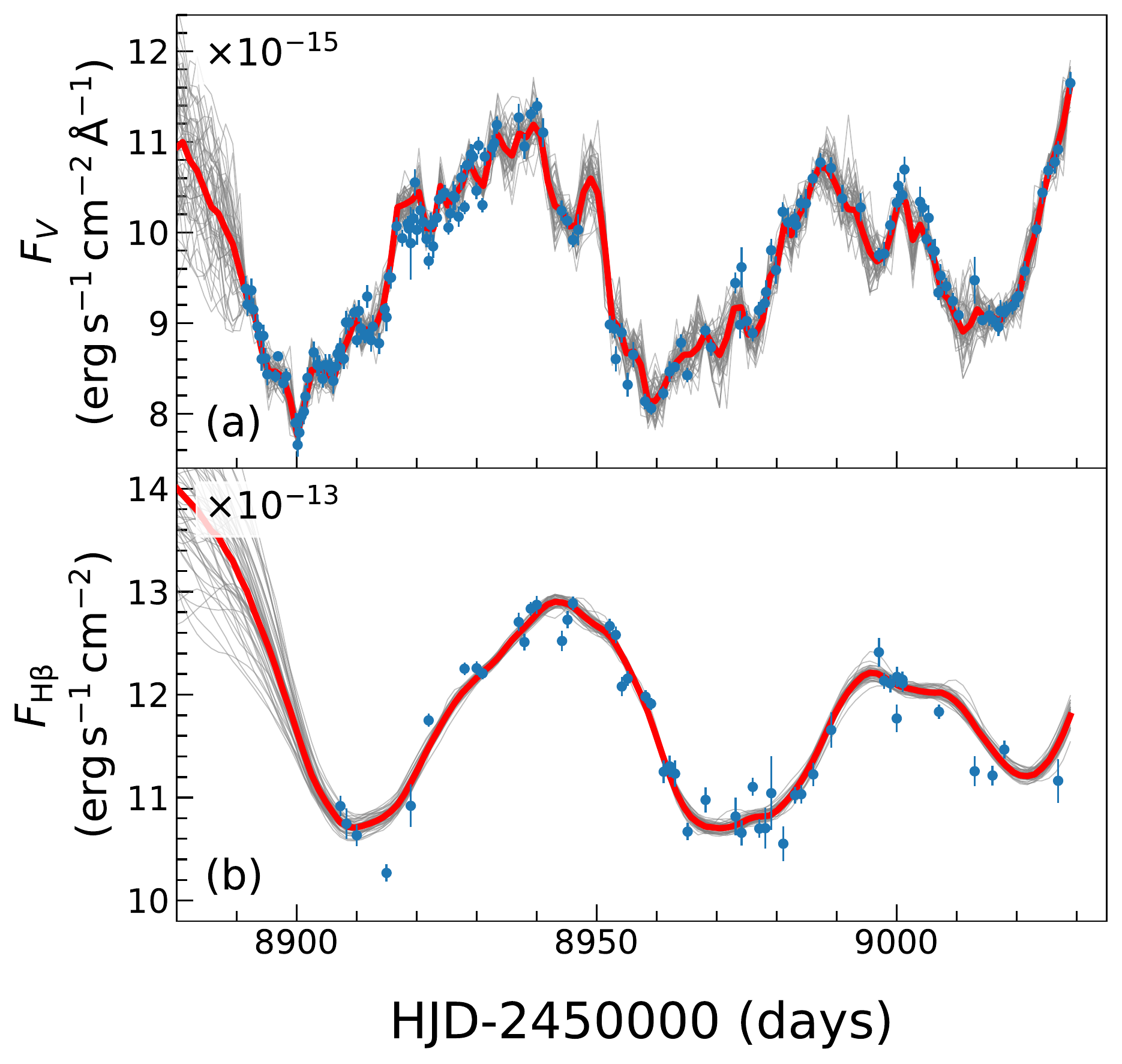}
\caption{Fitting results of (a) continuum and (b) \hb\ light curves.  We 
randomly selected 50 reconstructed continuum light curves and the reverberated 
\hb\ light curves from the posterior sample and display them in gray.  The 
median light curves are in red.  The data points are plotted in blue.}
\label{fig:lc}
\end{figure}

\begin{figure}
\centering
\includegraphics[width=0.45\textwidth]{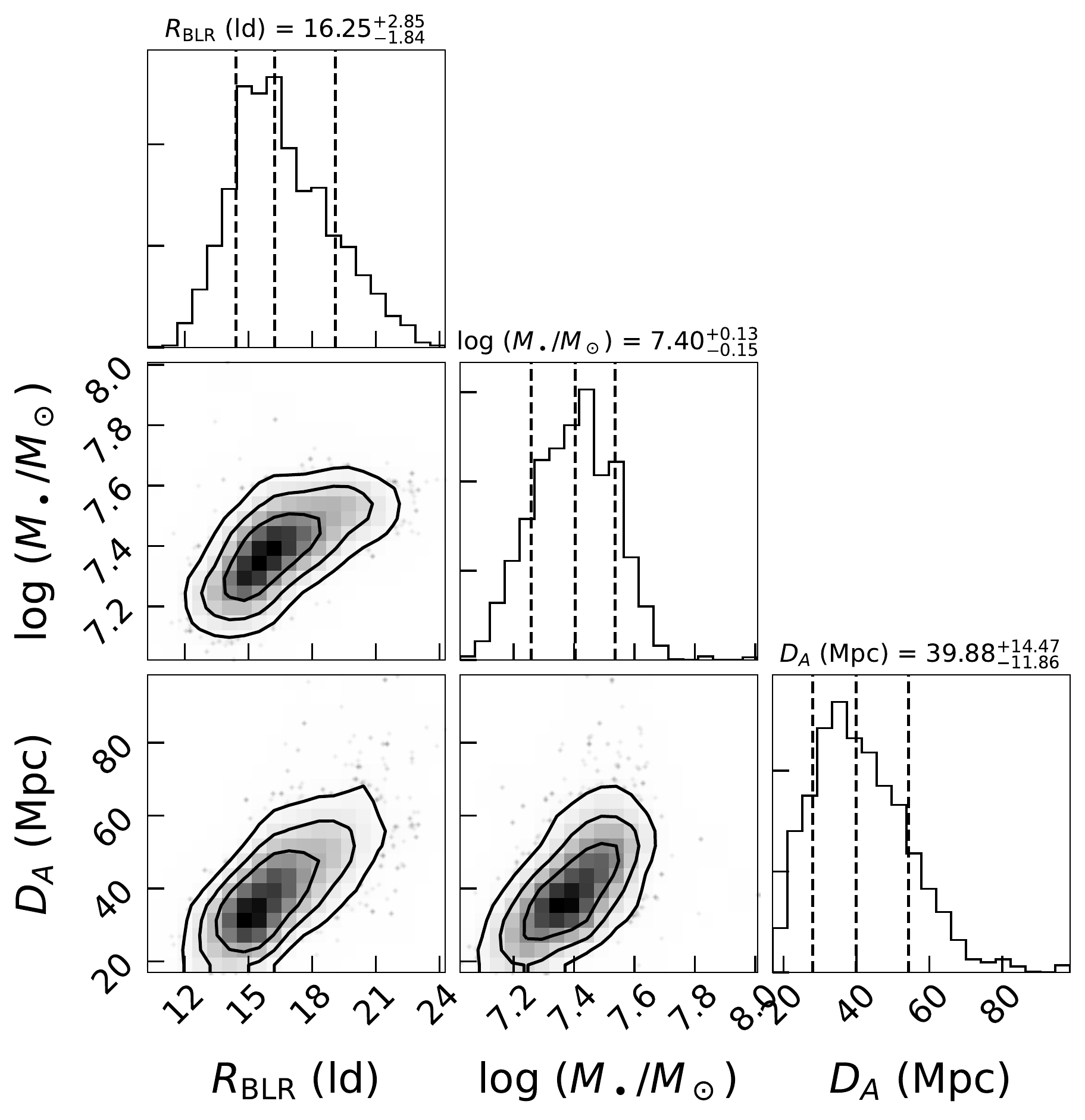}
\caption{Posterior probability distribution of the BLR radius, BH mass, and 
angular diameter distance of the joint analysis.  The dashed lines indicate 
the 16\%, 50\%, and 84\% percentiles of the of the posterior distributions. 
The contours indicate $1\sigma$, $1.5\sigma$, and $2\sigma$.}
\label{fig:cnk}
\end{figure}

\begin{figure}
\centering
\includegraphics[width=0.45\textwidth]{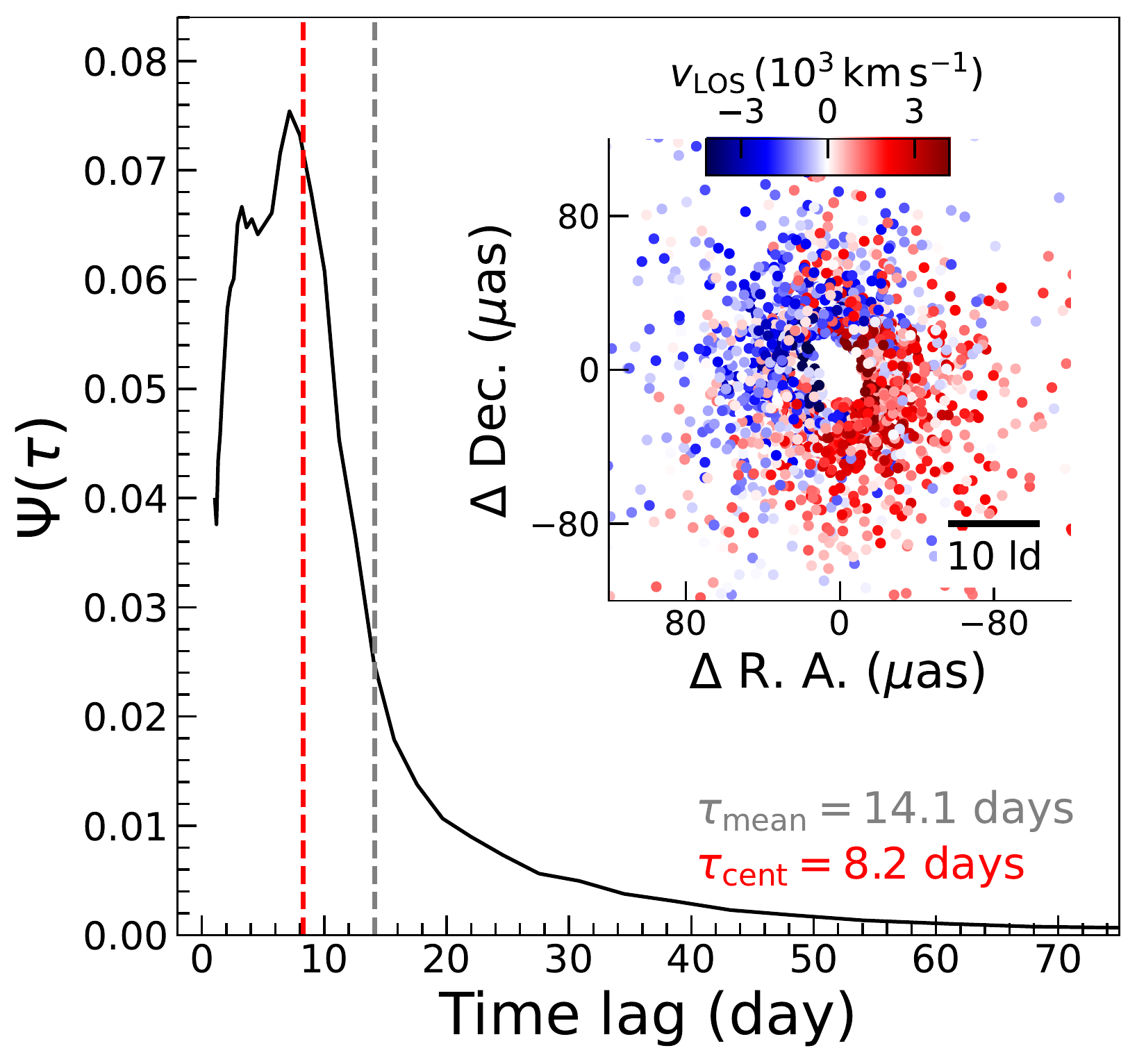}
\caption{Transfer function (Equation~(\ref{eq:lhb})) of the best-fit model 
peaks at the time lag close to $\tau_\mathrm{cent}$, measured from the 
CCF method.  But the mean lag ($\tau_\mathrm{mean}$) is longer because the 
transfer function has a long tail, which extends beyond the limit of the plot.
The inset on the right displays the cloud distribution of the best-fit BLR 
model.  The color code is the light-of-sight velocity.} 
\label{fig:tf}
\end{figure}

The line profile and differential phase signal based on the inferred median 
parameters are plotted in Figure~\ref{fig:phs}.  In this figure, the differential 
phase is averaged for UT4--UT2, UT4--UT1, and UT3--UT1.  The continuum phase 
(Section~\ref{sec:data}) is subtracted based on the median offset from the data 
of each baseline before the average.  The data averaged phase displays moderate 
asymmetry, the excess in data points with respect to the best-fit model seen on 
the blue side of the ``S'' shape, which can be explained by the uncertainty of 
those offsets.  The reconstructed continuum light curves and the reverberated 
\hb\ light curves are shown with the observed data in Figure~\ref{fig:lc}.  The 
detailed features of the continuum light curve are captured by the model, while 
the \hb\ light curve is reasonably well fitted.  We infer a correlation timescale 
$\tau_\mathrm{d} \approx 87$~day, slightly longer than the center of the prior 
distribution.  The mean time lag of the BLR model, based on the median inferred 
parameters, is $14.1_{-1.9}^{+2.3}$~day, which is close to the light-weighted 
radius of the BLR, $16.2_{-3.5}^{+5.4}$~ld (Figure~\ref{fig:cnk}).  However, the 
centroid of the cross correlation function (CCF) of the model continuum and 
H$\beta$ light curves is $\tau_\mathrm{cent}=8.2$~day.  Following 
\cite{Bentz2021ApJ}, we derive $\tau_\mathrm{cent}$ as the first moment of the 
CCF above 0.8 of the peak of the CCF \citep{Koratkar1991ApJS}.  Our 
$\tau_\mathrm{cent}$ is slightly lower than, but within 2-$\sigma$ to, that 
reported by \cite{Bentz2021ApJ}.  This is consistent with the conclusion in 
\citetalias{GCn3783}: The measured time lag from the CCF underestimates the 
BLR radius of NGC~3783 because the BLR radial distribution is strongly 
heavy-tailed (large $\beta$; Figure~\ref{fig:tf}).  Nevertheless, detailed 
comparisons of BLR structures of \brg\ and \hb\ lines, including the 
velocity-resolved RM, are needed in the future.  We discuss the fitting results 
considering an outer truncation of the BLR as well as the effect of nonlinear 
response of the continuum light curve in Appendix~\ref{apd:fit}.  The primary 
model introduced above is preferred to interpret the data.  While no 
disagreement is found statistically significant by introducing more physical 
constraints to the model, we caution that the inferred distance may be biased by 
the BLR model assumptions.

In order to assess the reliability of the uncertainties on the derived 
parameters, we considered the fitting process itself as well as exploring how 
sensitive they are to the light curve.  It is possible to adjust the temperature 
($T \geq 1$), by which the logarithmic likelihood is divided, with 
\texttt{CDNest} in order to enlarge the uncertainty of the data when the model 
is not flexible enough to fit the data \citep{Li2018zenodo,Brewer2011}.  We are 
always able to find the clear peak of the posterior weights as a function of the 
prior volume with $T=1$, indicating that the fitting has properly converged.  
The fitting results with $T>1$ are consistent with those with $T=1$, although 
the uncertainties increase.  Therefore, we always report the fitting results 
obtained using $T=1$.  The measured uncertainty of the \hb\ light curve is 
typically $\sim 1\%$ of the flux, which is lower than the typical uncertainty 
based on the intercalibration using the \OIII\ doublet \citep{Bentz2021ApJ}.  We 
fit the data with the \hb\ light curve uncertainties increased to be 2\% of the 
flux if they are smaller than the latter, and find that the results are fully 
consistent with those using the 
measured \hb\ uncertainties.  The uncertainty of \mbh\ reaches 0.2~dex, while 
that of $D_A$ barely increases as it is dominated by the phase data.  In order 
to further test whether the joint analysis is sensitive to the light curve 
measurement, we measured the \hb\ light curve by decomposing the spectra with a 
relatively wide wavelength range (see Appendix~\ref{apd:decomp} for more 
details).  Using the decomposition method, we measured the light curve with an 
approximatively 20\% lower variation amplitude.  We are able to obtain consistent fitting 
results once the nonlinear responsivity is taken into account in the joint 
analysis.

\section{Distance and peculiar velocity of NGC~3783}
\label{sec:dist}

Our joint analysis infers that the angular diameter distance of NGC~3783 is 
$39.9_{-11.9}^{+14.5}$~Mpc.  NGC~3783 has an observed redshift of 0.009730 and 
heliocentric velocity $v_h \approx 2917\,\kms$ \citep{Theureau1998AAS}.  
In this section, we derive the peculiar velocity of NGC~3783, compare our 
measured distance with other direct and indirect methods, and discuss the 
uncertainty of our measurements in detail.

\subsection{Peculiar velocity of NGC~3783} 
\label{sec:pec}

To correct for the motion of the Sun in the Milky Way and the peculiar motion of 
the Milky Way, we calculate the velocity of NGC~3783 in the frame of the Local 
Sheet \citep{Tully2008ApJ,Kourkchi2020AJ},
\begin{align}
v_\mathrm{ls} &= v_h - 26 \cos l \cos b + 317 \sin l \cos b - 8 \sin b 
\label{eq:vls} \\
              &= 2628\,\kms, \nonumber
\end{align}
where $(l, b)$ are the Galactic longitude and latitude.  The Local Sheet 
reference frame is a variant of the Local Group rest frame 
\citep{Yahil1977ApJ,Karachentsev1996AJ}.  \cite{Tully2008ApJ} advocated that the 
Local Sheet is preferable to the Local Group because it is more stable.
For comparison, NGC~3783 has a Local Group velocity 2627~\kms, according to 
NASA/IPAC Extragalactic Database (NED) velocity calculator.  The redshift 
corrected for the peculiar motion of the observing frame is 
$z_\mathrm{ls} = v_\mathrm{ls}/c \approx 0.008767$, 
which may still deviate from the cosmological redshift in the Hubble flow due 
to the peculiar motion of NGC~3783.  Assuming 
$(\Omega_m, \Omega_\Lambda, H_0)=(0.3, 0.7, 70)$, we can infer a cosmological 
redshift $\bar{z} = 0.0094_{-0.0028}^{+0.0035}$ according to the measured $D_A$. 
The peculiar velocity can be calculated \citep{Davis2014MNRAS},
\begin{equation}
v_\mathrm{pec} = c \frac{z_\mathrm{ls} - \bar{z}}{1 + \bar{z}} 
= -195_{-1029}^{+843}\,\kms.
\end{equation}
This is very close to, albeit with large uncertainties, the peculiar velocity 
estimated with the Cosmicflows-3 distance--velocity calculator (see below), 
$-182\,\kms$.  We also estimate the peculiar velocity using the 6dF 
galaxy redshift survey peculiar velocity map \citep{Springob2014MNRAS}.  We find 
11 galaxies in the 6dF peculiar velocity catalog within $8h^{-1}$~Mpc centered 
on the position of NGC~3783.  We averaged their peculiar velocities and 
uncertainties with equal weight.  The estimated peculiar velocity of NGC~3783, 
$-158 \pm 43\,\kms$, is consistent with the other estimates above.\footnote{We 
adopt the recession velocity, 3197~\kms\ in the CMB frame, of the galaxy group 
that comprises NGC~3783 \citep{Kourkchi2017ApJ} in order to obtain the 3-D 
supergalactic coordinate of NGC~3783.  We adopt the typical smoothing scale, 
$8h^{-1}$~Mpc, to find the galaxies sharing the same large-scale structure with 
NGC~3783.  We find it makes little difference if we weight the peculiar velocity 
according to their separation or not, so we average them with equal weight for 
simplicity.  The peculiar velocity has been converted from the CMB frame 
\citep{Planck2020AAa} to the Local Sheet frame in order to be compared with 
other estimates.}

\begin{table}
\centering
\renewcommand{\arraystretch}{1.5}
\begin{tabular}{c | c  }
\hline\hline
Distance (Mpc)         & Method description                          \\ \hline
$39.9_{-11.9}^{+14.5}$ & SARM (this work)                            \\
$49.8 \pm 19.6$        & Tully-Fisher relation \citep{Robinson2021ApJ}  \\
$38.5 \pm 14.2$        & Tully-Fisher relation \citep{Tully1988}     \\
35.1                   & Cosmicflows-3 \citep[EDD;][]{Kourkchi2020AJ} \\
37.9                   & NED (Virgo + GA + Shapley)                  \\
42.1                   & Galaxy group \citep{Kourkchi2017ApJ}        \\
\hline\hline
\end{tabular}
\caption{Distance of NGC~3783 measured with different methods.  The SARM 
and Tully-Fisher relation are the direct methods, while the other methods are 
indirect.  The distance uncertainties of the indirect methods are discussed in 
the text whenever they can be estimated.
}
\label{tab:dist}
\end{table}

\subsection{Comparison with other distance measures} 
\label{sec:dcp}

Besides our measured $D_A$, the distance of NGC~3783 can be measured 
``directly'' with the Tully-Fisher relation \citep{Tully1977AA} on 
the one hand and estimated ``indirectly'' with various methods relying on 
the large-scale structure and velocity field of the local universe on the other.  
Strictly speaking, these methods, as discussed below, provide the luminosity 
distance.  Since the redshift of NGC~3783 is too low for there to be a significant 
difference between angular diameter distance and luminosity distance ($<2\%$), 
we do not distinguish the two in the following discussion.  The direct distance 
measurements and indirect estimates that we discuss below are summarized in 
Table~\ref{tab:dist}.

The distance of NGC~3783 has been measured as $38.5 \pm 14.2$~Mpc by 
\cite{Tully1988} using the Tully-Fisher relation.  This measurement was based on 
the magnitude of the galaxy without removing the nuclear emission from the BH 
accretion.  This will bias the distance toward a lower value.  
\cite{Robinson2021ApJ} recently derived the distance of NGC~3783 to be 
$49.8 \pm 19.6$~Mpc.  They derived the maximum rotational velocity measured from 
the newly observed \hi\ spectrum and measured the multiband optical/NIR 
magnitudes of the galaxy after carefully decomposing the nuclear emission.  
Although they adopt a 20\% uncertainty of the distance throughout the entire 
sample, the best-estimate distance of NGC~3783 is based on the HST photometry as 
the value quoted above.  Nevertheless, \cite{Robinson2021ApJ} emphasized that the 
distance of NGC~3783 is quite uncertain mainly because the galaxy is rather 
face-on, leading to a large uncertainty of the maximum rotational velocity.
The strong bar of NGC~3783 may also influence the rotation velocity of the galaxy 
\citep[e.g.,][]{Randriamampandry2015MNRAS} and affect the distance measured 
from the Tully-Fisher relation.

Based on the Cosmicflows-3 catalog \citep[CF3,][]{Tully2016AJ} of the distance of 
$\sim 18000$ galaxies, \cite{Graziani2019MNRAS} reconstruct the smoothed 
peculiar velocity field, for the first time, up to $z \approx 0.05$ using a 
linear density field model.  They assume a fiducial cosmology 
$(\Omega_m, \Omega_\Lambda, H_0)=(0.3, 0.7, 75)$ in the modeling while 
considering deviations of $H_0$ from the fiducial value, so that the 
reconstructed velocity field does not depend on the assumed $H_0$.  They find 
the cosmic expansion is consistent with their fiducial $H_0$ 
\citep[see also][]{Tully2016AJ}.  We use the CF3 distance--velocity calculator 
provided by the Extragalactic Distance Database 
(EDD)\footnote{\url{http://edd.ifa.hawaii.edu/}.} 
to estimate the luminosity distance of NGC~3783 based on its $v_\mathrm{ls}$ 
corrected for cosmological effects \citep{Davis2014MNRAS,Kourkchi2020AJ}, 
\begin{align}
v^c_\mathrm{ls} &= v_\mathrm{ls} \left[1 + \frac{1}{2} (1 - q_0) z_\mathrm{ls} - 
\frac{1}{6} (2 - q_0 - 3 q_0^2) z_\mathrm{ls}^2 \right], \label{eq:vc}\\
q_0 &= \frac{1}{2} (\Omega_m - 2 \Omega_\Lambda).
\end{align}
Adopting $\Omega_m=0.3$ and $\Omega_\Lambda=0.7$, we find $q_0=-0.55$ and 
NGC~3783 has $v^c_\mathrm{ls}=2646~\kms$.\footnote{It is worth noting that 
$v^c_\mathrm{ls}$ does not depend on $H_0$ \citep{Davis2014MNRAS}.  And the 
correction with Equation~(\ref{eq:vc}) is $<1\%$ as $z_\mathrm{ls}$ is very 
small.}  The corresponding distance of NGC~3783 is 35.1~Mpc according to the CF3 
calculator.  It is not straightforward to estimate the uncertainty associated with 
this method.  We expect that the principal source of uncertainty comes from 
the peculiar velocity of the galaxy due to the nonlinear effects that cannot be 
described by the linear model of \cite{Graziani2019MNRAS}.  They approximate 
the nonlinear effects in the model with a single parameter of nonlinear velocity 
dispersion $\sigma_\mathrm{NL}$.  Taking the face value of 
$\sigma_\mathrm{NL} \approx 280\,\kms$ from their Bayesian inference, the linear 
$v^c_\mathrm{ls}$ of NGC~3783 is in the range 2366--2926~\kms.  The CF3 calculator 
yields distances ranging from 30.1 to 41.8~Mpc.  Therefore, we estimate 
the uncertainty of the distance from the CF3 calculator is at least 
6~Mpc.\footnote{Another calculator, NAM, based on nonlinear model of galaxies 
within 38~Mpc (or 2850~\kms) is provided by EDD.  However, NGC~3783 is 
just on the upper boundary of the NAM calculator.  It yields 
$D_L=37.5$~Mpc fully consistent with that of the CF3 calculator, although the NAM 
result is likely much more uncertain as the nonlinear model is poorly 
constrained at the edge of application \citep{Kourkchi2020AJ}.  We therefore 
prefer the result from the CF3 calculator.}

For completeness, NED provides the estimates of galaxy distances based on 
the multiattractor model by \cite{Mould2000ApJ}.  We obtain a Hubble flow 
velocity 2638~\kms\ after correcting for the infall due to Virgo cluster, Great 
Attractor, and Shapley supercluster.  Taking our fiducial $H_0=70\,\uhc$, the 
distance of NGC~3783 is 37.9~Mpc, very close to the result of CF3.  NGC~3783 is 
found in a group of 9 galaxies according to an updated nearby galaxy catalog by 
\cite{Kourkchi2017ApJ}.  The weighted distance using the 2 galaxies with 
measured distance from Cosmicflows-3 is $42.1 \pm 5.9$~Mpc.  Although with 
considerably large uncertainty, this estimate provides an indication on the 
absolute uncertainty of the indirect approach.

\begin{figure}
\centering
\includegraphics[width=0.45\textwidth]{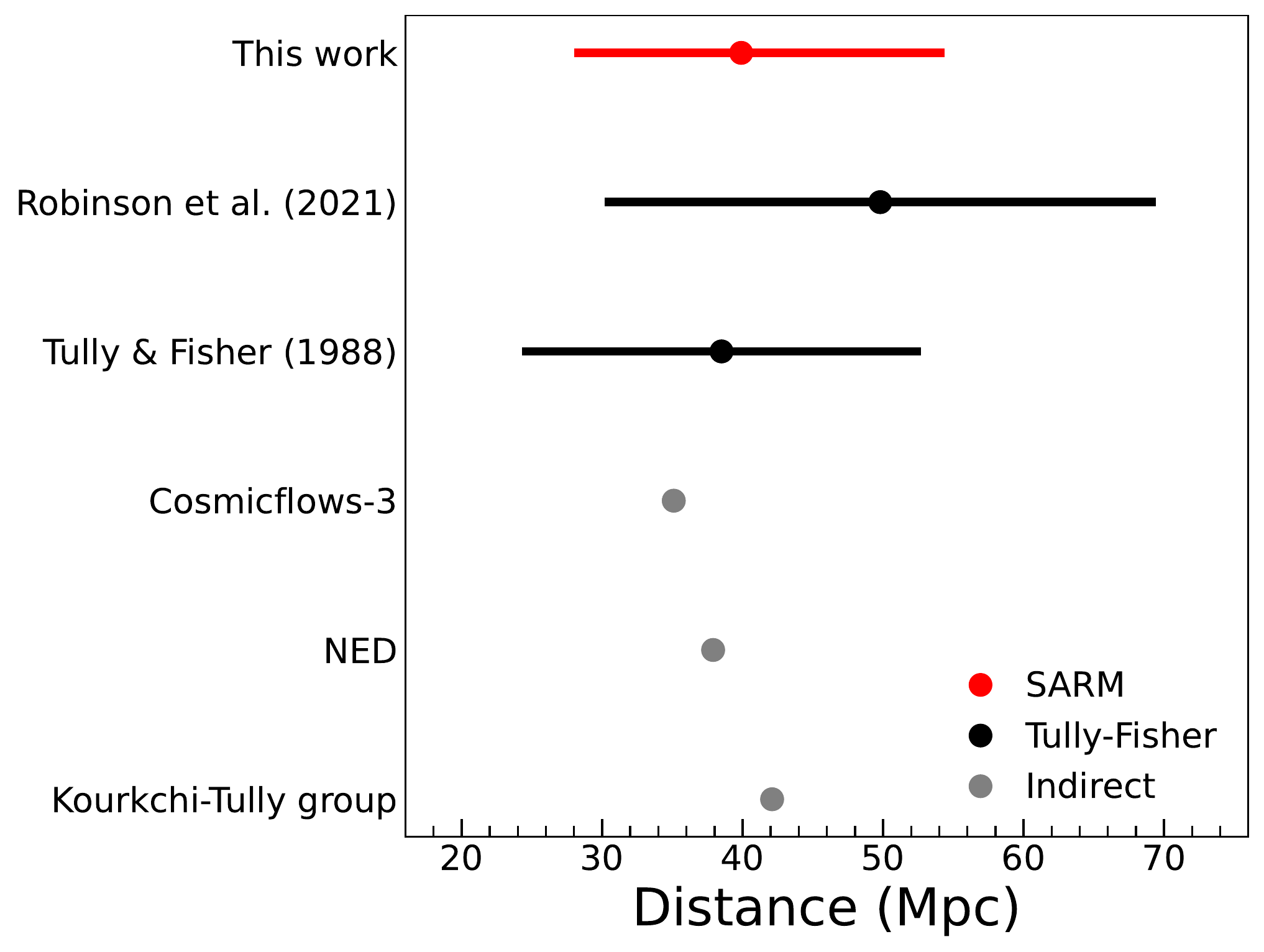}
\caption{Comparison of the distances of NGC~3783 measured with the methods 
listed in Table~\ref{tab:dist}.  The 1-$\sigma$ uncertainties of direct 
measurements are plotted, while those of indirect measurements are not likely 
smaller than the former.  We discuss the uncertainties of the indirect 
measurements in the text whenever they can be estimated.
}
\label{fig:dist}
\end{figure}

As displayed in Figure~\ref{fig:dist}, our newly measured distance is fully 
consistent with the results based on other direct and indirect methods, among 
which the RMS scatter is $\sim 5$~Mpc.  The uncertainty of our measurement is 
comparable to those of the Tully-Fisher relation.

\subsection{Statistical and systematic uncertainties}
\label{sec:sys}

The spectral resolution of GRAVITY is moderate and noise of adjacent channels is 
likely correlated.  This may lead to an underestimate of the inferred 
uncertainty.  To gauge the effect of the correlated noise, we perform the 
analysis in two ways, (1) using only half of GRAVITY differential phase spectra 
(every second channel) and (2) rebinning across every two channels in the 
differential phase spectra.  The inferred parameter uncertainties are typically 
$\lesssim 10\%$ higher than those reported in Table~\ref{tab:par}.  
Therefore, we conclude that the effect of the correlated phase noise is 
moderate.\footnote{In contrast to the 2-$\sigma$ credible interval used in 
\citetalias{GCIRAS} and \citetalias{GCn3783}, we report the 1-$\sigma$ 
uncertainty level throughout this paper for the simplicity of making fair 
comparisons to the distance measurements with other methods.  The posterior 
distributions of most of the key parameters have profiles close to Gaussian, so 
the choice of the credible interval does not affect our conclusions.}

The relative uncertainty of our measured distance is about 33\%, which is the 
combination of the uncertainty of the linear radius ($R_\mathrm{BLR}$) and the 
angular radius ($\Theta_\mathrm{BLR}$) 
of the BLR.  We find $R_\mathrm{BLR}$ has a relatively small uncertainty of about 14\%, 
and is mainly constrained by the RM data.  In contrast, $\Theta_\mathrm{BLR}$ has about 25\% 
uncertainty, mainly due to the relatively large uncertainty of the differential 
phase.  The differential phase measurements have about 0.1\degree\ uncertainty 
per baseline, so the relative uncertainty of the $\sim 0.2\degree$ phase signal 
on three baselines is about 30\%.  Therefore, we find 
$\delta_{D_A} \approx \sqrt{\delta_{R_\mathrm{BLR}}^2 + \delta_{\phi}^2}$, 
where $\delta_{D_A}$, $\delta_{R_\mathrm{BLR}}$, and $\delta_{\phi}$ are the 
fractional uncertainties of the distance, linear radius of the BLR, and the 
differential phase combining 3 baselines.  This is overall consistent with the 
conclusion of \cite{Songsheng2021ApJS} based on tests with mock data.

A primary concern in terms of systematic uncertainty is that the BLR structure of 
different lines measured by RM and GRAVITY may be different.  A 
detailed comparison of \hb\ and \brg\ BLR structures for NGC~3783 is out of the 
scope of this work and will be studied in a separate paper where the 
velocity-resolved RM modeling will be used.  Alternatively, the broad line 
profile contains the information of the BLR structure.  Following 
\cite{Wang2020NatAs}, we can estimate the relative size difference between the 
\hb\ and \brg\ BLRs using $(q-1)/(q+1)$, where 
$q=\left(\mathrm{FWHM_{H\beta}/FWHM_{Br\gamma}}\right)^2$.
We measure $\mathrm{FWHM}_\mathrm{Br\gamma} \approx 3680\,\kms$ from our 
SINFONI spectrum \citepalias{GCn3783}.  As introduced in more detail in 
Appendix~\ref{apd:decomp}, we fit globally the \hb\ complex of 50 
RM spectra individually, including the narrow and broad components of \hb\ and 
\heiia, \OIIIc, and \feii\ emission, as well as the power-law continuum 
\citep[e.g.,][]{Barth2015ApJS,Hu2015ApJ}.
Accounting for the instrumental broadening \citep{Bentz2021ApJ}, we find 
the FWHM of \hb\ is stable over the RM campaign and 
$\mathrm{FWHM}_\mathrm{H\beta} = 4341 \pm 197\,\kms$.  Our result is slightly 
smaller than the $\mathrm{FWHM}_\mathrm{H\beta}$ of the mean spectrum reported 
by \cite{Bentz2021ApJ} mainly due to our different approaches to fit the 
continuum.  Therefore, the size difference between the \hb\ and \brg\ BLRs can be 
about $\sim 15\%$, which is comparable to the 13\% RMS of different distance 
measurements (Section~\ref{sec:dcp}).  Calculations with photoionization models 
will also provide useful insights \citep{Zhang2021ApJ}, although special attention 
needs to be paid to the difficulty of reproducing the observed flux ratios of 
Hydrogen lines in CLOUDY \citep{Netzer2020MNRAS}.  RM observations of the same 
line as that observed with GRAVITY can avoid this problem.

The short time lag of NGC~3783 makes it much more straightforward to 
compare the BLR of different lines from RM and GRAVITY than that of 3C~273.  
The $\sim$100 day time lag of 3C~273 requires the campaign to be at least 
several years.  The continuum light curve suffers from a further complication caused by 
an overall long-term trend \citep{Li2020ApJ}.  In fact, the newly measured lag is about 
2~times smaller than the result of early RM campaigns \citep{Kaspi2000ApJ}.
The dynamical timescale of the BLR \citep{Peterson1993PASP} in NGC~3783 is 
$t_\mathrm{dyn}\approx R_\mathrm{BLR} / v_\mathrm{FWHM}\approx 3$~years.  The 
BLR structure likely varies on a timescale $\gtrsim t_\mathrm{dyn}$ (e.g., 
\citealt{Peterson1993PASP,Peterson2004ApJ,Lu2016ApJ}).  This explains why we 
find that the \brg\ spectra in GRAVITY measurements from 2018 to 2020 show 
overall consistent line profiles, and that these are also consistent with the 
profile from the SINFONI observation in 2019.  Although it is not possible to 
entirely exclude it, we do not expect significant size variation in the BLR 
between GRAVITY and RM observations.  Meanwhile, we emphasize that 
quasi-simultaneous observation of GRAVITY and RM within $t_\mathrm{dyn}$ is 
necessary to avoid issues associated with the variability of the BLR structure. 

\subsection{\bf Toward a new estimate of the Hubble constant} \label{sec:h0}

With a measurement of $D_A$, one natural step forward is to estimate the 
Hubble constant. 
We derive $H_0 = v^c_\mathrm{ls}/\left[D_A (1+z_\mathrm{ls})^2\right] = 
65_{-17}^{+28}\,\kms\,\mathrm{Mpc}^{-1}$ from the joint analysis of NGC~3783.  
The peculiar velocity of NGC 3783 is not included in this estimate.  Our 
estimate (Sec.~\ref{sec:pec}) suggests the additional systematic uncertainty 
introduced by the peculiar velocity is $\lesssim 10\%$.  At this level, the 
differential phase errors dominate the uncertainty of $H_0$. With upgrades to 
GRAVITY including adaptive optics with laser guide star and wide 
angle off-axis phase referencing, GRAVITY+ will allow us to observe at least 
several hundred suitable Type~1 AGN targets up to redshift 2--3, with \paa, 
Pa$\beta$, Pa$\gamma$, H$\alpha$, and \hb\ redshifted into $K$ band.  These 
lines are much stronger than \brg\ with respect to the continuum.  Therefore, 
we expect much less statistical uncertainty (e.g., $\lesssim 10\%$) on $D_A$ 
with future observations.  Assuming the same sensitivity level as the current 
GRAVITY performance, \cite{Songsheng2021ApJS} show that it might in principle be 
possible to obtain a precision measurement of $H_0$ using a large sample of 
differential phase and RM measurements. In practice, more effort would be 
necessary to address the potential bias due to the properties of different 
hydrogen lines and the assumptions of the BLR model.

\section{Conclusions}\label{sec:con}

NGC~3783 is the second AGN observed by both GRAVITY interferometry and RM 
campaigns.  We fitted both data sets simultaneously with a SARM joint analysis.  
The inferred model parameters are fully consistent with the previous study using 
only GRAVITY data, and the uncertainties of key model parameters are 
significantly reduced.  In particular, the inferred BH mass is 
$2.54_{-0.72}^{+0.90}\times 10^7\,M_\odot$ (68\% credible interval).  For this 
parameter, the uncertainty is a factor of two smaller than both GRAVITY-only 
inference and RM 
measurements using the integrated \hb\ light curve (dominated by the 0.3~dex 
calibration uncertainty of the virial factor).  We also confirm the previous 
finding of \citetalias{GCn3783} that the BLR is highly concentrated with an 
extended tail of clouds out to large radius, which leads to an apparent 
discrepancy between the mean radius of the BLR and the measured time lag.

With the joint analysis, we are able to constrain the angular diameter distance 
of NGC~3783 to be $39.9^{+14.5}_{-11.9}$~Mpc.  Because NGC~3783 is in the nearby 
Universe, this distance can be compared to other independent direct and indirect 
measurements based on the Tully-Fisher relation, galaxy flow models, and its 
galaxy group.  The BLR-based geometric distance is fully consistent with 
these other results.  The dominant uncertainty for the distance comes from the 
differential phase, which has a relative uncertainty of about 30\%, while the 
relative uncertainty from RM data is about 14\%.  

With the ongoing upgrade of GRAVITY, we expect to observe many AGNs with 
improved sensitivity.  Our analysis indicates that the uncertainties of BLR 
model parameters, such as $\Theta_\mathrm{BLR}$ and \mbh, will be reduced as the 
phase uncertainty decreases.   Being able to substantially reduce the 
statistical uncertainty of $H_0$ looks promising by both improving the precision 
of individual measurements and averaging the measurements over many targets.  
In this context, it will be important to better understand the systematic 
uncertainties associated with the SARM method.  Further, measuring the same 
broad line in $K$ band with both RM and GRAVITY will be important for future study.

\begin{acknowledgements}

We thank the referees for their careful reading of the manuscript and 
their suggestions that have helped to improve the clarity of it.
We thank Bradley Peterson for his helpful comments on this paper.
J.S. would like to thank the important help from Yan-Rong Li and Yu-Yang 
Songsheng to implement and test the joint analysis model.  He also thanks the 
useful discussion from Jian-Min Wang and the IHEP group. 
J.D. was supported in part by NSF grant AST 1909711 and an Alfred P. Sloan Research Fellowship.  
M.C.B. gratefully acknowledges support from the NSF through grant AST-2009230 to Georgia State University.  
A.A. and P.G. were supported by Funda\c{c}\~{a}o para a Ci\^{e}ncia e a Tecnologia, with grants reference UIDB/00099/2020, PTDC/FIS-AST/7002/2020 and SFRH/BSAB/142940/2018. 
SH acknowledges support from the European Research Council via Starting Grant ERC-StG-677117 DUST-IN-THE-WIND. 
JSB acknowledges the full support from the UNAM PAPIIT project IA 101220.  
P.O.P. acknowledges financial support from the CNRS « Programme national des hautes énergies » and from the french space agency CNES.
This research has made use of the NASA/IPAC Extragalactic Database (NED), which 
is operated by the Jet Propulsion Laboratory, California Institute of 
Technology, under contract with the National Aeronautics and Space 
Administration. This research made use of 
\textsc{Astropy},\footnote{http://www.astropy.org} a community-developed core 
Python package for Astronomy \citep{Astropy:2013ek,Astropy:2018dk}, 
\textsc{numpy} \citep{vanderWalt:2011we}, \textsc{scipy} \citep{Jones:2001ch}, 
and \textsc{matplotlib} \citep{Hunter:2007}
\end{acknowledgements}

\bibliography{blr}{}
\bibliographystyle{aa}

\appendix

\section{The model of the continuum light curve}
\label{apd:lc}

The observed continuum light curve can be described by 
\citep{Rybicki1992ApJ,Li2018ApJ},
\begin{equation}
\vec{l_c} = \vec{s} + \vec{E} q + \vec{n_c},
\end{equation}
where the vector $\vec{s}$ is the variable component of the light curve with 
mean zero, $\vec{E} q$ represents the mean flux of the light curve, and 
$\vec{n_c}$ is the noise of the observation.  $\vec{E}$ is a vector of unity 
with the same length of the observed light curve and $q$ is the mean of the 
light curve.  Assuming that the light curve is a multivariate Gaussian with 
 covariance described by Equation~(\ref{eq:drw}) and the measurement 
uncertainties are Gaussian and uncorrelated, we can interpolate and extrapolate 
the light curve with a uniformly and densely sampled continuum light curve 
model.  The observed light curve is sampled in a sparse uneven time series, 
$\vec{t_1}$, meanwhile the model light curve is densely sampled in a uniform 
time series, $\vec{t_2}$.  The posterior mean and covariance of the variable 
component of the model light curve can be calculated,
\begin{align}
\hat{\vec{s}} &= (\vec{C_{11}}^{-1}\vec{C_{12}})^T (\vec{l_c} - \vec{E} q), \\
\vec{C_s} &= \vec{C_{22}} - (\vec{C_{11}}^{-1}\vec{C_{12}})^T \vec{C_{12}},
\end{align}
where $\vec{C_{11}}=S(\vec{t_1}, \vec{t_1}) + \vec{\sigma_c}^2 \vec{I}$, 
$\vec{C_{12}}=S(\vec{t_1}, \vec{t_2})$, and 
$\vec{C_{22}}=S(\vec{t_2}, \vec{t_2})$ are matrices calculated based on 
Equation~(\ref{eq:drw}), and superscript ``T'' denotes transposition.  We denote 
the observation uncertainties as $\vec{\sigma_c}$ and the identity matrix as 
$\vec{I}$. The mean and variance of the long-term average of the model light 
curve are,
\begin{align}
\hat{q} &= C_q \vec{E}^T \vec{C_{11}}^{-1} \vec{l_c}, \\
C_q &= (\vec{E}^T \vec{C_{11}}^{-1} \vec{E})^{-1}.
\end{align}
Therefore, the model continuum light curve is,
\begin{equation}
\vec{\Tilde{l}_c} = (\vec{L} \vec{x_s} + \hat{\vec{s}}) + \vec{E} (\sqrt{C_q} x_q + \hat{q}),
\end{equation}
where $\vec{L}$ is the lower triangular matrix of $\vec{C_s}$, so that 
$\vec{C_s}=\vec{L}\vec{L}^T$, $\vec{x_s}$ describes the deviation of the light 
curve variable component from its posterior mean values, and $x_q$ describes the 
deviation of the long-term mean from its posterior mean value.  We fit 
$\vec{x_s}$ and $x_q$ in the joint analysis as free parameters for the continuum 
light curve model.

\section{Model fitting results and further tests}
\label{apd:fit}

\begin{figure}
\centering
\includegraphics[width=0.45\textwidth]{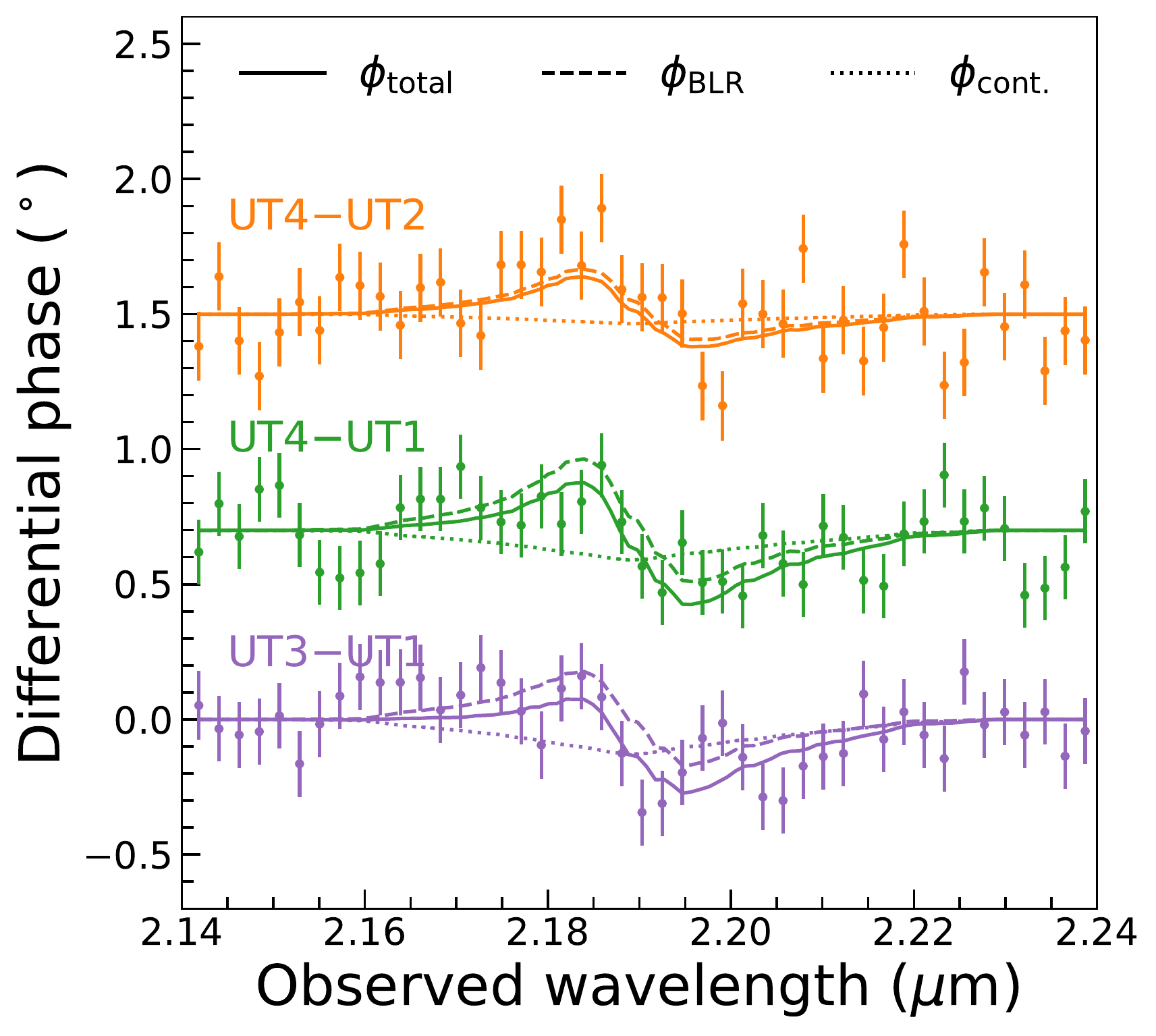}
\caption{Averaged differential phase data (points) and the best-fit models 
(lines) of the three baselines that show the strongest signal of the BLR 
component (dashed lines). The best-fit residual continuum phases are shown with 
dotted lines.}
\label{fig:phson}
\end{figure}

\begin{figure*}
\centering
\includegraphics[width=0.95\textwidth]{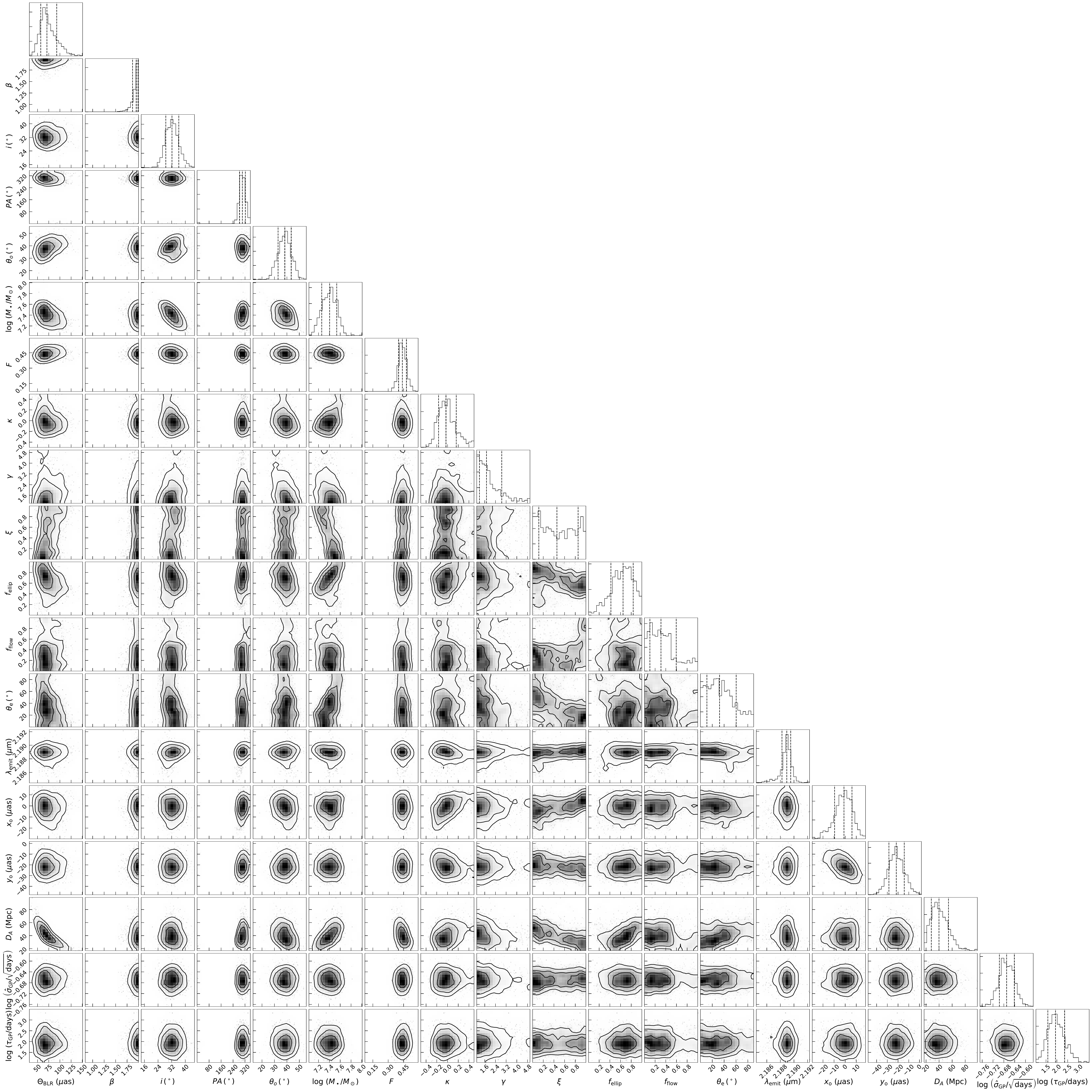}
\caption{Posterior distribution of physical parameters from the Bayesian 
joint analysis.}
\label{fig:corner}
\end{figure*}

\begin{table}
\centering
\renewcommand{\arraystretch}{1.5}
\begin{tabular}{l | c }
\hline\hline
                          Parameters    &  Model Inference              \\ \hline
$\Theta_\mathrm{BLR}\,(\mu\mathrm{as})$ & $71_{-13.8}^{+21.9}$          \\ 
$R_\mathrm{BLR}\,(\mathrm{ld})$         & $16.2_{-1.8}^{+2.8}$          \\ 
$\Theta_\mathrm{min}\,(\mu\mathrm{as})$ & $31_{-7}^{+11}$               \\ 
$\beta$                                 & $1.95_{-0.08}^{+0.04}$        \\
$\theta_\mathrm{o}\,(^\circ)$           & $38_{-5}^{+5}$                \\
$i\,(^\circ)$                           & $32_{-4}^{+4}$                \\
PA ($^\circ$ E of N)                    & $302_{-19}^{+20}$             \\ 
$\kappa$                                & $-0.03_{-0.14}^{+0.19}$       \\ 
$\gamma$                                & $1.7_{-0.5}^{+1.1}$           \\ 
$\xi$                                   & $0.5_{-0.3}^{+0.4}$           \\ 
Offset ($\mu$as)                        & $\left(-1.1_{-8.4}^{+7.2}, 
                                           -21.6_{-6.9}^{+7.4}\right)$  \\
$\log\,(\mbh/M_\odot)$                  & $7.40_{-0.15}^{+0.13}$        \\
$f_\mathrm{ellip}$                      & $0.65_{-0.23}^{+0.19}$        \\
$P$(inflow)                             & 0.79                          \\
$\theta_\mathrm{e}\,(^\circ)$           & $32_{-21}^{+28}$              \\
$\Delta v_\mathrm{BLR}\,(\mathrm{km\,s^{-1}})$
                                        & $328_{-97}^{+81}$             \\
$D_A$ (Mpc)                             & $39.9_{-11.9}^{+14.5}$        \\
$\log\,(\tau_\mathrm{d}/\mathrm{day})$
                                        & $1.94_{-0.35}^{+0.42}$        \\
$\log\,(\hat{\sigma}_\mathrm{d}/\sqrt{\mathrm{day}})$      
                                        & $-0.67_{-0.03}^{+0.03}$       \\ \hline
$\chi_\mathrm{r}^2$                     & 0.672                         \\ 
\hline\hline
\end{tabular}
\caption{Posterior median and 68\% credible intervals for the BLR model 
parameters.  The central offset of the BLR is in (R. A., Dec.).  $P$(inflow) is 
the probability of inflow where $f_\mathrm{flow}<0.5$ in the posterior sample.  
$\Delta v_\mathrm{BLR}$ is the difference between the velocity derived from the 
best-fit $\lambda_\mathrm{emit}$ and the systemic velocity based on the redshift. 
}
\label{tab:par}
\end{table}

As discussed in Section~\ref{sec:data}, we use the differential phase data after 
subtracting the continuum phase primarily due to the secondary continuum 
emission component according to the reconstructed image \citepalias{GCn3783}.  
In order to display the goodness of the fit, we display the differential phases 
of UT4--UT2, UT4--UT1, and UT3--UT1, averaged over the three epochs, in 
Figure~\ref{fig:phson} together with the best-fit models.  The phase signal is 
dominated by the BLR component while the residual continuum phase is moderate. 
The BLR phase signal is not significantly detected in the other three short baselines.

The posterior probability distributions of all of the physical parameters are 
shown in Figure~\ref{fig:corner}.  As discussed in Section~\ref{sec:infe}, the 
posterior distributions of the BLR model of the joint analysis show very good 
consistency with those derived using GRAVITY data alone \citepalias{GCn3783}.
The uncertainties of several key parameters, such as $\beta$, 
$\theta_\mathrm{o}$, $i$, and \mbh, are reduced by a factor of 
$\sim 2$, while those of the other parameters stay the same.  The reduction of 
uncertainties of the inferred parameters is also found in the joint analysis of 
3C~273 \citep{Wang2020NatAs}.  This indicates that RM data help to constrain the 
model in a consistent manner with GRAVITY data.  The $\beta$ parameter is closer 
to the upper boundary than GRAVITY only inference, reinforcing our previous 
finding that the BLR is heavy-tailed.  With GRAVITY data only, we found $\xi$ 
peaks at $\sim 1$ with a long tail to $\sim 0$, indicating that $\xi$ is not 
strongly constrained.  With the joint analysis, $\xi$ is even less constrained 
with a nearly flat posterior distribution (see Figure~\ref{fig:corner}).  This 
means that the fitting is not sensitive to the midplane obscuration, mainly 
because most of the BLR clouds ($65\%$) are in elliptical orbits rather 
than radial orbits.  The best-fit model favors moderate inflow, which is 
consistent with \cite{Bentz2021ApJ}.

\subsection{Constraining the BLR outer radius}
\label{apd:rmax}

We further test whether the fitting results are significantly different when 
we constrain the BLR inner and outer radii based on previous RM of high-ionization 
lines and NIR continuum.  Following the discussion of \citetalias{GCn3783}, we 
cast additional prior information that $R_\mathrm{min}=4$~ld and 
$R_\mathrm{out}=80$~ld.  The key results are shown in Table~\ref{tab:prmax}, 
where $D_A$ is reduced by about 30\%.  The angular size of the BLR is almost the 
same as our primary model without additional constraints, as it is directly 
constrained by the GRAVITY phase data.  At the same time, $R_\mathrm{BLR}$ is reduced by about 
30\% due to the hard boundary $R_\mathrm{out}=80$~ld of the cloud distribution.  
Given the observed angular size of the BLR, a smaller $D_A$ is required in order 
to incorporate a smaller BLR as required by the prior.  In addition, the inferred 
$\beta$ is around 1, meaning that the cloud distribution is now exponential rather than 
heavy-tailed.  This is because $R_\mathrm{out}$ limits the extent of the 
cloud distribution.  The peak of the model differential phase is correspondingly 
reduced, although the model fits the data almost as well as our primary 
model given the data uncertainty.  All of the other inferred model parameters 
are consistent with our primary model results.  Nevertheless, the 
radius-constrained model is not preferred over our primary model because the 
inferred $D_A$ would be the smallest of the measurements discussed in 
Section~\ref{sec:dist}.  Quantitatively, it is about 1.5-$\sigma$ lower than the 
other measurements, which cautions against the additional constraints to the model.  
While $R_\mathrm{out}=80$~ld is a reasonable choice based on NIR continuum RM 
analyses \citep{Glass1992MNRAS,Lira2011MNRAS}, the contribution of NIR emission 
from the accretion disk may bias the time lag to a small value 
\citep{Lira2011MNRAS}.  Nevertheless, the joint analysis may also be biased by 
the assumptions of the BLR model.  More comprehensive comparisons will be 
performed in future paper that considers different model assumptions, e.g., 
different cloud radial distributions and BLR size constraints.

\begin{table*}
\centering
\renewcommand{\arraystretch}{1.5}
\begin{tabular}{l | r  r  r }
\hline\hline
            \multirow{2}{*}{Parameters} &  \multicolumn{3}{c}{Model Inference with}                           \\ 
                                        & (1) Boundary           & (2) Nonlinearity      & (3) New light curve    \\ \hline
$\Theta_\mathrm{BLR}\,(\mu\mathrm{as})$ & $71_{-14}^{+19}$       & $70_{-16}^{+27}$       & $74_{-16}^{+21}$       \\ 
$R_\mathrm{BLR}\,(\mathrm{ld})$         & $12.7_{-1.6}^{+1.8}$   & $18.9_{-4.9}^{+4.6}$   & $16.4_{-5.6}^{+6.5}$   \\ 
$D_A$ (Mpc)                             & $30.9_{-7.0}^{+8.6}$   & $43.6_{-17.8}^{+24.7}$ & $36.1_{-12.6}^{+21.6}$ \\
$\log\,(\mbh/M_\odot)$                  & $7.51_{-0.17}^{+0.31}$ & $7.40_{-0.19}^{+0.14}$ & $7.50_{-0.19}^{+0.23}$ \\
\hline\hline
\end{tabular}
\caption{Same as Table~\ref{tab:pk}, but showing the fitting results: 
(1) when  $R_\mathrm{min}=4$~ld and $R_\mathrm{max}=80$~ld are required (Appendix~\ref{apd:rmax}); 
(2) when the nonlinear response is considered (Appendix~\ref{apd:nonl}); 
(3) when we use the light curve measured by the decomposition method (Appendix~\ref{apd:decomp}).
}
\label{tab:prmax}
\end{table*}

\subsection{The nonlinear response of RM}
\label{apd:nonl}

The BLR reverberates the Hydrogen ionizing (UV) photon variability from the 
accretion disk.  The optical (i.e., $V$ band) light curve is usually used as an 
approximation to the UV light curve.  However, their correlation is not 
necessarily linear.  Following \cite[][see also \citealt{Li2013ApJ}]{Li2018ApJ}, 
we revise Equation~(\ref{eq:lhb}) into
\begin{equation}\label{eq:nonl}
\Tilde{l}_\mathrm{H\beta}(t) = A \int \Psi(\tau) \, l_c^{1+\delta_c}(t - \tau) \, d\tau,
\end{equation}
where $\delta_c$ captures the nonlinearity of the response.  The model 
inference is conducted in the same way for the other parameters as discussed in 
Section~\ref{sec:join}, with the uniform prior of $\delta_c$ between -1 and 3 
\citep{Li2018ApJ}.  We find the inferred model parameters are entirely 
consistent with our primary fitting results (Table~\ref{tab:par}).  Some of the 
key parameters are listed in Table~\ref{tab:prmax}.  Meanwhile, the uncertainty 
of $R_\mathrm{BLR}$ increases by about 70\%, resulting in an increase to the 
uncertainty of $D_A$ by the same amount.  This mainly originates from the 
degeneracy between $\delta_c$ and the scaling factor $A$ 
(Equation~(\ref{eq:nonl})).  Optimizing the sampling method will reduce the 
uncertainties of some parameters due to the degeneracy.  However, this is beyond 
the scope of the current paper.  On the other hand, the fitting of the light 
curve is not substantially improved.  Therefore, we opt not to include the 
nonlinear effect of the RM response in the modeling.

\subsection{Measuring the light curve using the decomposition method}
\label{apd:decomp}

Here we test whether the joint analysis is sensitive to the method measurement 
the light curve.  \cite{Bentz2021ApJ} measured the \hb\ flux by fitting a 
local linear continuum underneath the \hb\ line.  This method has been widely 
used and has been shown to be effective in producing accurate light curves for 
strong emission lines, such as \hb\ 
\citep[e.g.,][]{Peterson1998ApJ,Kaspi2000ApJ,Grier2012ApJ}.  Recent studies have 
developed methods to decompose the emission lines by simultaneously fitting 
various physical components to the spectra over a wide wavelength range 
\citep{Barth2015ApJS,Hu2015ApJ,Shen2016ApJ}.  We therefore conducted a joint fit 
using the \hb\ light curve measured with the decomposition method in order to 
test whether the joint analysis is sensitive to the measurement of the light 
curve.  Following the approach of \cite{Barth2015ApJS}, our model consists of 
the AGN power-law continuum, \OIII\ doublet, broad and narrow \hb\ and \heii\ 
lines, and the \feii\ blended lines.  We find the stellar continuum, reddening, 
and \hei\ lines at 4471, 4922, and 5016~\AA\ are not necessary to achieve a good 
fit so, to improve the stability of the decomposition of different spectra, we 
do not include these components.  The broad and narrow emission lines are all 
fitted with a fourth-order Gauss–Hermite function \citep{vanderMarel1993ApJ}, 
whose centroid, dispersion, $h_3$, $h_4$, and amplitude are the parameters in 
the fitting.  The profiles of the other narrow lines (the dispersion, $h_3$, and 
$h_4$) are tied to that of \OIIIb.  The ratio of \OIIIc\ is fixed to 1/3.  The 
rest-frame wavelength separation of the two lines is fixed to 47.9~\AA.  The 
\feii\ emission lines are modeled with the template from 
\cite{Boroson1992ApJS}.  We adjust the redshift, dispersion, and amplitude of 
the template in the fitting.  We fit the spectra at rest-frame 4500--5500~\AA, 
masking the small wavelength range of \NI\ lines at 5199 and 5201~\AA.  We 
subtract all of the best-fit components except the narrow and broad \hb\ line, 
and the \hb\ flux is integrated over rest-frame 4800--4940~\AA, in order to be 
consistent with \cite{Bentz2021ApJ}.  To estimate the uncertainty of the light 
curve, we perturb the data according to the measurement uncertainty of the 
spectra and redo the measurement.  The final uncertainty is the quadrature sum 
of the statistical uncertainty from the bootstrapping and 1\% of the integrated 
flux to make sure that the resulting uncertainty is not too small.

\begin{figure}
\centering
\includegraphics[width=0.45\textwidth]{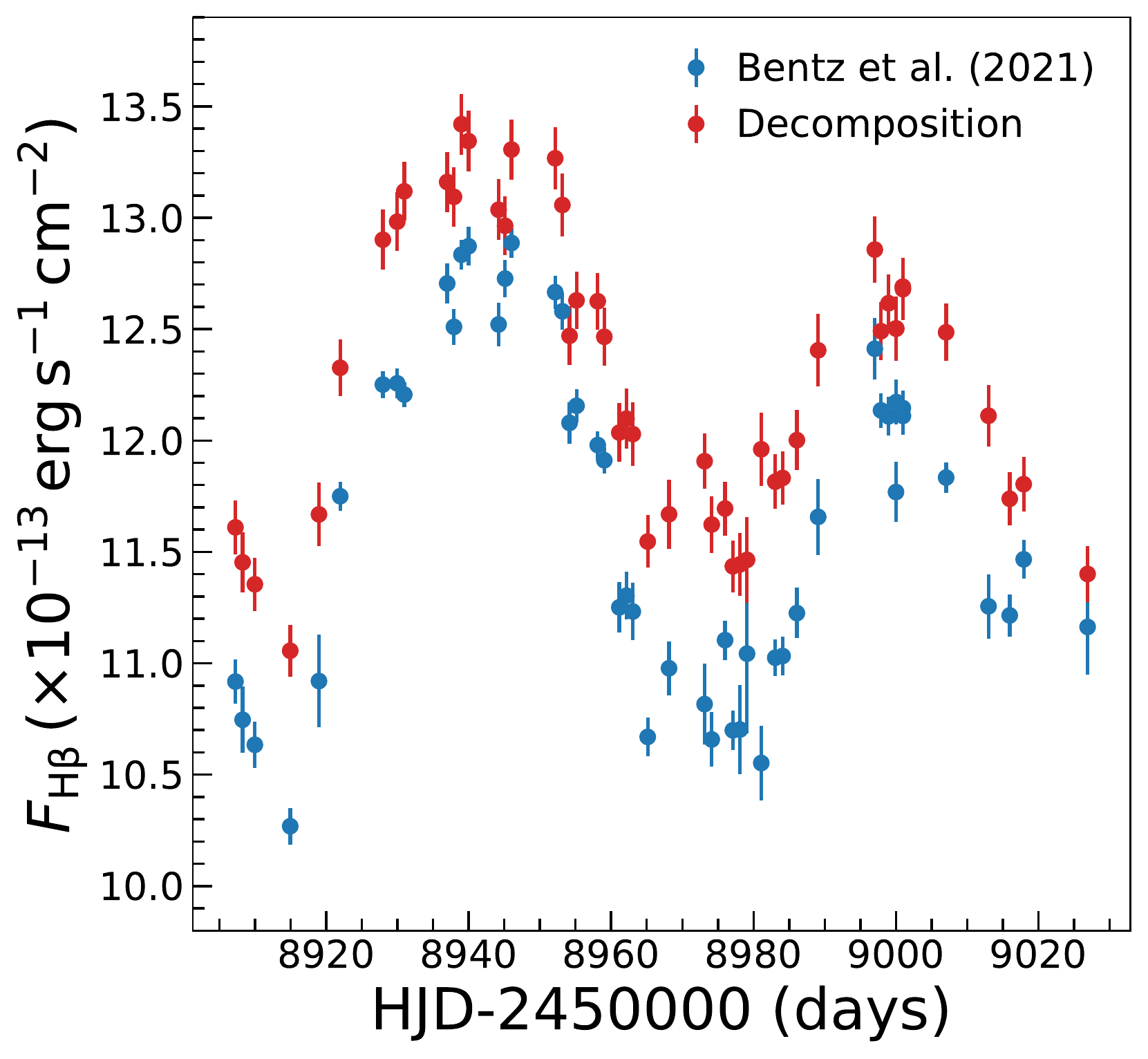}
\caption{Comparison of the \hb\ light curves measured by 
\citet[][in blue]{Bentz2021ApJ} and the decomposition method (in red, see the 
text).}
\label{fig:lccmp}
\end{figure}

The light curve measured by the decomposition method has a very similar shape to 
that from \cite{Bentz2021ApJ} as shown in Figure~\ref{fig:lccmp}.  The former is 
about 5\% higher than the latter, probably because the best-fit continuum from 
the decomposition method is slightly lower than that from the local continuum 
underneath the \hb\ line.  Using the method introduced in 
Section~\ref{sec:join}, we are able to infer model parameters consistent with 
those reported in Table~\ref{tab:par} within 1-$\sigma$.  However, the 
peak-to-valley variation amplitude of the new light curve is reduced by about 
20\%.  Correspondingly, the new inference prefers about 30--50\% larger BLR 
radius, inclination angle, and $D_A$ in order to reduce the variation amplitude 
of \hb\ light curve.  The model inference is slightly biased in this way.  To 
compensate for this, we need to take the nonlinearity of the \hb\ response 
(Equation~\ref{eq:nonl}) into account.  This yields inferred parameters almost 
exactly the same as those reported in Table~\ref{tab:par} with the additional 
parameter $\delta_c \approx -0.2$.  The key parameters are listed in 
Table~\ref{tab:prmax}.  The responsivity of \hb\ is physically expected to be 
$\lesssim 1$ (i.e., $\delta_c < 0$) based on photoionization models 
\citep[e.g.,][]{Gaskell1986ApJ,Korista2004ApJ} as well as observations of e.g., 
NGC~5548 \citep{Goad2014MNRAS}.  We emphasize that the observational effects, 
such as using the $V$-band continuum as a proxy for the ionization continuum and 
spectral decomposition over a wide wavelength range, may also introduce a 
nonlinear effect.  The change of the light curve variation amplitude does not 
influence much the time lag of the RM.  However, our test shows that the joint 
analysis may be biased if the nonlinear effect is not properly taken into 
account.

\end{document}